\documentclass[aps,floats]{revtex4-2}
\usepackage{amsmath,amssymb}
\usepackage{graphicx,epsfig}
\usepackage[greek,english]{babel}
\usepackage{bbold}

\begin{document}
\bibliographystyle {plain}

\pdfoutput=1
\def\oppropto{\mathop{\propto}} 
\def\opsimeq{\mathop{\simeq}}
\def\opoverderline{\mathop{\overline}}
\def\operarrow{\mathop{\longrightarrow}}
\def\opsim{\mathop{\sim}}

\def\opmin{\mathop{\min}} 
\def\opmax{\mathop{\max}} 
\def\oplim{\mathop{\lim}}

\title{ A supersymmetric quantum perspective on the explicit large deviations 
for reversible Markov jump processes, with applications to pure and random spin chains   } 


\author{C\'ecile Monthus}
\affiliation{Universit\'e Paris-Saclay, CNRS, CEA, Institut de Physique Th\'eorique, 91191 Gif-sur-Yvette, France}


\begin{abstract}

The large deviations at various levels that are explicit for Markov jump processes satisfying detailed-balance are revisited in terms of the supersymmetric quantum Hamiltonian $H$ that can be obtained from the Markov generator via a similarity transformation. We first focus on the large deviations at level 2 for the empirical density ${\hat p}(C) $ of the configurations $C$ seen during a trajectory over the large time-window $[0,T]$ and rewrite the explicit Donsker-Varadhan rate function as the matrix element $I^{[2]}[{\hat p}(.) ] = \langle \sqrt{ {\hat p} } \vert H \vert \sqrt{{\hat p}} \rangle $ involving the square-root ket $\vert \sqrt{{\hat p}} \rangle $. [The analog formula is also discussed for reversible diffusion processes as a comparison.] We then consider the explicit rate functions at higher levels, in particular for the joint probability of the empirical density ${\hat p}(C) $ and the empirical local activities ${\hat a}(C,C') $ characterizing the density of jumps between two configurations $(C,C')$. Finally, the explicit rate function for the joint probability of the empirical density ${\hat p}(C) $ and of the empirical total activity ${\hat A} $ that represents the total density of jumps of a long trajectory is written in terms of the two matrix elements $ \langle \sqrt{ {\hat p} } \vert H \vert \sqrt{{\hat p}} \rangle$ and $\langle \sqrt{ {\hat p} } \vert H^{off} \vert \sqrt{{\hat p}} \rangle $, where $H^{off} $ represents the off-diagonal part of the supersymmetric Hamiltonian $H$. This general framework is then applied to pure or random spin chains with single-spin-flip or two-spin-flip transition rates, where the supersymmetric Hamiltonian $H$ correspond to quantum spin chains with local interactions involving Pauli matrices of two or three neighboring sites. It is then useful to introduce the quantum density matrix ${\hat \rho} = \vert \sqrt{{\hat p}} \rangle \langle \sqrt{ {\hat p} } \vert $  associated to the empirical density ${\hat p}(.) $ in order to rewrite the various rate functions in terms of reduced density matrices involving only two or three neighboring sites.

\end{abstract}

\maketitle


\section{ Introduction   }

The language of large deviations is very useful to unify many areas of statistical physics
 (see the reviews \cite{oono,ellis,review_touchette} and references therein)
 and has become essential to formulate the properties of non-equilibrium stochastic dynamics
(see the reviews with various perspectives \cite{derrida-lecture,harris_Schu,searles,harris,mft,sollich_review,lazarescu_companion,lazarescu_generic,jack_review}, 
the PhD Theses \cite{fortelle_thesis,vivien_thesis,chetrite_thesis,wynants_thesis,chabane_thesis,duBuisson_thesis} 
 and the Habilitation Thesis \cite{chetrite_HDR}).
In particular, the large deviations properties of time-averages over a long Markov trajectory
have attracted a lot of interest in many different contexts 
 \cite{peliti,derrida-lecture,sollich_review,lazarescu_companion,lazarescu_generic,jack_review,vivien_thesis,lecomte_chaotic,lecomte_thermo,lecomte_formalism,lecomte_glass,kristina1,kristina2,JackSollich2010,simon1,simon2,tailleur,simon3,Gunter1,Gunter2,Gunter3,Gunter4,chetrite_canonical,chetrite_conditioned,chetrite_optimal,chetrite_HDR,touchette_circle,touchette_langevin,touchette_occ,touchette_occupation,garrahan_lecture,Vivo,c_ring,c_detailed,chemical,derrida-conditioned,derrida-ring,bertin-conditioned,touchette-reflected,touchette-reflectedbis,c_lyapunov,previousquantum2.5doob,quantum2.5doob,quantum2.5dooblong,c_ruelle,lapolla,c_east,chabane,us_gyrator,duBuisson_gyrator,c_largedevpearson,sollich_MFtrap} with the construction of the corresponding Doob's conditioned processes.
Since the rate functions for these time-averaged observables are not always explicit,
a key idea has been to consider the large deviations at higher levels 
in order to obtain explicit rate functions for arbitrary Markov processes.
Since the level 2 concerning the distribution of the empirical density 
that has been much studied since the pioneering works of Donsker and Varadhan \cite{DonskerV}
is not explicit for arbitrary non-equilibrium Markov processes,
an important progress has been the formulation
of the level 2.5 concerning the joint distribution of the empirical density and of the empirical flows,
where rate functions are always explicit, either for discrete-time Markov chains 
\cite{Hollander,fortelle_thesis,fortelle_chain,review_touchette,Polettini,c_largedevdisorder,c_reset,c_inference,carugno,c_microcanoEnsembles,c_chaotic},
for continuous-time Markov jump processes 
\cite{fortelle_thesis,fortelle_jump,maes_canonical,maes_onandbeyond,wynants_thesis,chetrite_formal,BFG1,BFG2,chetrite_HDR,c_ring,c_interactions,c_open,barato_periodic,chetrite_periodic,c_reset,c_inference,c_LargeDevAbsorbing,c_microcanoEnsembles,c_susyboundarydriven,c_Inverse},
for diffusion processes 
\cite{wynants_thesis,maes_diffusion,chetrite_formal,engel,chetrite_HDR,c_lyapunov,c_inference,c_susyboundarydriven,c_missing,c_SmallNoise}, 
as well as for jump-diffusion or jump-drift processes \cite{c_reset,c_runandtumble,c_jumpdrift,c_SkewDB}.
Then the large deviations properties at any lower level can be formally obtained from this explicit
 level 2.5 via the operation called 'contraction' :
one needs to optimize the level 2.5 over the empirical observables that one wishes to integrate over,
 in order to see if the rate function for the remaining empirical observables can be written explicitly. 
 
 The goal of the present paper is to revisit the various levels that can be written
explicitly for any Markov jump process satisfying detailed-balance,
 from the perspective of the supersymmetric quantum Hamiltonian $H$ 
 that can be obtained from the Markov generator via a similarity transformation.
 
 The paper is organized as follows. 
In section \ref{sec_revSusy}, we introduce the notations for reversible Markov jump generators and their
associated supersymmetric quantum Hamiltonians.
In section \ref{sec_level2} devoted to the large deviations at level 2 concerning the empirical density ${\hat p}(C) $, 
we explain why the supersymmetric quantum Hamiltonian $H$ is useful to rewrite
the explicit Donsker-Varadhan rate function at level 2.
In section \ref{sec_activities}, we recall the explicit level 2.5 concerning the joint distribution of the empirical density ${\hat p}(C) $ and of the empirical transitions between configurations in order to discuss various intermediate levels between 2.5 and 2 whose rate functions are still explicit for reversible Markov jump processes.
This general framework is then applied in section \ref{sec_spins} to pure or random spin chains with local transition rates, where the supersymmetric Hamiltonian $H$ correspond to quantum spin chains with local interactions involving Pauli matrices of two or three neighboring sites.
Our conclusions are summarized in section \ref{sec_conclusions}.
In Appendix \ref{app_diffusion}, we describe how the large deviations properties of reversible diffusion processes
involve supersymmetric quantum Hamiltonians in continuous space
in order to stress the similarities and the differences with Markov jump processes
described in the main text.


\section{ Reversible Markov jump processes and their supersymmetric Hamiltonians    }

\label{sec_revSusy}

In this section, we first introduce the notations for arbitrary Markov jump generators 
and then focus on the specific properties associated to detailed-balance,
in particular on their associated supersymmetric quantum Hamiltonians.

\subsection{ Markov jump process of generator $w$ converging towards the steady state $p_*(C)$ }

 The dynamics for the probability $p_t( C ) $ to be on configuration $C$ at time $t$
satisfies the master equation
\begin{eqnarray}
 \partial_t p_t( C )    =   \sum_{C' }  w(C,C') p_t(C')
\label{master}
\end{eqnarray}
where the off-diagonal $C \ne C'$ positive matrix element $w(C,C') \geq 0$    
represents the transition rate from $C'$ to $C $,
while the negative diagonal elements $w(C,C) \leq 0 $ 
are determined in terms of the off-diagonal elements
by the conservation of probability 
\begin{eqnarray}
w(C,C) \equiv  - \sum_{C' \ne C} w(C',C) 
\label{wdiag}
\end{eqnarray}
The steady state $p_*(C) $ of Eq. \ref{master} satisfies
\begin{eqnarray}
0 =     \sum_{C' }  w(C,C') p_*(C')  
\label{mastereqst}
\end{eqnarray}
Eqs \ref{wdiag} and \ref{mastereqst} mean that 
 the highest eigenvalue of the Markov matrix $w(.,.)$ vanishes,
with the trivial positive left eigenvector 
\begin{eqnarray}
 l_0(C)=1
\label{markovleftj}
\end{eqnarray}
and the positive right eigenvector $r_0(C)$ given by the steady state
\begin{eqnarray}
 r_0(C)=p_*(C) 
\label{markovrightj}
\end{eqnarray}

Note that the present paper is written with the vocabulary and the notations of the physics literature that differ from the mathematic literature: in particular the matrix $w(.,.)$ governing the master Eq. \ref{master} will be called the generator,
and its matrix element $w(C,C') \geq 0$ represents the transition rate from the right configuration $C'$ towards the left configuration $C $ in order to have the same notations as in quantum mechanics.
 
 
 \subsection{ Specific properties in the presence of detailed-balance}

\subsubsection{ Parametrization of the transition rates $ w(C',C)$ satisfying detailed-balance }

The master equation of Eq. \ref{master}
 can be rewritten as a continuity equation
\begin{eqnarray}
 \partial_t p_t( C )    =    \sum_{C' \ne C }  j_t(C,C')
\label{mastercontinuity}
\end{eqnarray}
where the current $ j_t(C,C')$ from $C'$ to $C$ at time $t$ involves the two rates $w(C,C') $ and $w(C',C) $
\begin{eqnarray}
 j_t( C,C' )    = w(C,C') p_t (C') - w(C',C) p_t(C)  = - j_t( C',C )
\label{mastercurrent}
\end{eqnarray}

The dynamics converging towards the steady state $p_*(C)$
satisfies detailed-balance if all the steady currents $  j_*( C,C' )$ vanish  
\begin{eqnarray}
0=  j_*( C,C' )   = w(C,C') p_* (C') - w(C',C) p_*(C)
\label{detailed}
\end{eqnarray}
It is then convenient to parametrize the transitions rates as
\begin{eqnarray}
w(C,C') = D(C,C') \sqrt{ \frac{ p_* (C)}{p_* (C')} }
\label{detailedratio}
\end{eqnarray}
in terms of the steady state $p_*(.)$ and in terms of the symmetric matrix with positive elements
\begin{eqnarray}
 D(C',C) =  D(C,C') \geq 0
\label{dcc}
\end{eqnarray}
associated to the links $C' \ne C$.

The Markov dynamics is said to be reversible when the detailed-balance of Eq. \ref{detailed} is satisfied, 
while it is said to be irreversible when there are some non-vanishing steady currents $  j_*( C,C' ) \ne 0$.


\subsubsection{ Similarity transformation of the generator $w(.,.)$ into a quantum Hamiltonian $H(.,.)$ }

As explained in textbooks \cite{gardiner,vankampen,risken} and in specific applications to various models (see for instance \cite{glauber,Felderhof,siggia,kimball,peschel,jpb_antoine,pierre,texier,us_eigenvaluemethod,Castelnovo,c_pearson,c_boundarydriven}), generators of reversible Markov processes are related to quantum hermitian Hamiltonians
via similarity transformations.
For Markov jump processes,
 the standard change of variable 
\begin{eqnarray}
p_t ( C ) \equiv \sqrt{p_* (C) }  \psi_t ( C )
\label{relationPpsi}
\end{eqnarray}
transforms the master equation of Eq. \ref{master} for $p_t ( C ) $
into the Euclidean Schr\"odinger equation for $\psi_t ( C )$ 
\begin{eqnarray}
\partial_t \psi_t ( C ) = - \sum_{C'} H(C,C') \psi_t ( C' )
\label{Hquantum}
\end{eqnarray}
where the quantum Hamiltonian $H$ is obtained from the generator $w$ via the similarity transformation
\begin{eqnarray}
H(C,C') = - \frac{1}{ \sqrt{p_* (C) }} w(C,C') \sqrt{p_* (C') }
\label{Hsimilarity}
\end{eqnarray}
The off-diagonal matrix elements of the Hamiltonian  are simply given by the opposite
of the positive symmetric function $D(C,C')$ introduced in the parametrization of Eq. \ref{detailedratio}
\begin{eqnarray}
H(C,C') = - D(C,C')= - D(C',C)  = H(C',C)  \ \ \ \text{ for } \ \  C' \ne C
\label{offdiag}
\end{eqnarray}
while the diagonal matrix elements reduce to the opposite of the diagonal elements $w(C,C) $ of Eq. \ref{wdiag}
\begin{eqnarray}
H(C,C)  = - w(C,C) 
=\sum_{C' \ne C } w(C',C)  =  \sum_{ C' \ne C }  D(C',C)  \sqrt{ \frac{ p_* (C')}{p_* (C)} }
\label{diag}
\end{eqnarray}
So the Hamiltonian $H$ is a real symmetric matrix with negative off-diagonal matrix elements 
and positive diagonal matrix elements.
Via Eq. \ref{relationPpsi}, the steady state $p_* ( C ) $ corresponds to the positive ground state
\begin{eqnarray}
  \psi_0 (C) = \sqrt{p_* (C) }
\label{psi0}
\end{eqnarray}
of the quantum Hamiltonian $H$ associated to zero-energy
\begin{eqnarray}
 H \vert \psi_0 \rangle =0 
\label{HGS}
\end{eqnarray}
If there are ${\cal N}$ configurations $C$, the spectral decomposition 
of the Hamiltonian 
\begin{eqnarray}
 H = \sum_{n=1}^{{\cal N}-1} E_n \vert \psi_n \rangle  \langle \psi_n \vert
\label{Hspectral}
\end{eqnarray}
involves the other $({\cal N}-1)$ positive energies $E_n>0$ and their associated real eigenvectors $\psi_n(C)$
satisfying the orthonomalization and closure relations
\begin{eqnarray}
 \delta_{n,m} && = \langle \psi_n  \vert \psi_m \rangle = \sum_C \psi_n(C) \psi_m(C) 
 \nonumber \\
 \mathbf{1} && = \sum_{n=0}^{{\cal N}-1}  \vert \psi_n \rangle  \langle \psi_n \vert
\label{orthopsi}
\end{eqnarray}
The propagator of the Markov process can be then rewritten via the similarity transformation of Eq. \ref{Hsimilarity} 
in terms of the quantum propagator with its spectral decomposition
\begin{eqnarray}
P_t( C\vert C_0 ) \equiv \langle C \vert e^{w t} \vert C_0 \rangle 
 = \frac{\psi_0(C) }{\psi_0(C_0)} \langle C \vert e^{-  H t} \vert C_0 \rangle
&& = \frac{\psi_0(C) }{\psi_0(C_0)} \left[ \psi_0 (C) \psi_0 (C_0) + \sum_{n=1}^{{\cal N}-1} e^{- t E_n } \psi_n (C) \psi_n (C_0)\right] 
\nonumber \\
&& =  \psi^2_0 (C) 
+ \sum_{n=1}^{{\cal N}-1} e^{- t E_n } \left[ \psi_n (C) \psi_0(C) \right] \frac{  \psi_n (C_0)}{\psi_0(C_0)}
\label{propagator}
\end{eqnarray}
yielding that the $({\cal N}-1)$ positive energies $E_n>0$ of $H$ represent the relaxation spectrum of the Markov process
towards its steady state $p_* (C) = \psi^2_0 (C)  $.


\subsubsection{ Supersymmetric properties of the quantum Hamiltonian $H$ }

\label{subsec_susy}

The quantum Hamiltonians obtained from reversible Markov processes
via similarity transformations are not arbitrary but can be rewritten as supersymmetric Hamiltonians $H=Q^{\dagger} Q$
in terms of another operator $Q$ and its adjoint $Q^{\dagger}$.
This supersymmetric factorization that is very standard for diffusion processes in continuous space (see for instance  \cite{jpb_antoine,pierre,texier,c_lyapunov,us_gyrator,c_pearson,c_boundarydriven}) is recalled around Eq. \ref{hsusy} of Appendix \ref{app_diffusion}.
For Markov jump processes, this supersymmetric factorization
has been less studied except in one dimension \cite{c_boundarydriven} where it displays very special properties,
so it seems useful in the present subsection to describe the general case of an arbitrary configuration space.

The very specific structure of the matrix elements of Eqs \ref{offdiag} and \ref{diag}
yields that the quantum Hamiltonian $H$ can be rewritten as a sum over the links 
after choosing some order $C'>C$
 \begin{eqnarray}
H \equiv \sum_C \sum_{C'} \vert C \rangle H(C,C') \langle C' \vert  =  \sum_{C_1} \sum_{C_2>C_1} H^{[C_2 , C_1]}
\label{Hsum}
\end{eqnarray}
where the Hamiltonian $H^{[C_2 , C_1]} $ associated to the oriented link $[C_2 , C_1] $ 
\begin{eqnarray}
H^{[C_2 , C_1]}  && =   D(C_2,C_1)  
\left[ \sqrt{\frac{\psi_0 (C_1)}{\psi_0 (C_2)} } \vert C_2  \rangle  - \sqrt{\frac{\psi_0 (C_2)}{\psi_0 (C_1)} } \vert C_1  \rangle    \right] 
 \left[ \sqrt{\frac{\psi_0 (C_1)}{\psi_0 (C_2)} } \langle C_2  \vert  - \sqrt{\frac{\psi_0 (C_2)}{\psi_0 (C_1)} } \langle C_1  \vert      \right] 
\nonumber \\
&&  \equiv  \vert q^{[C_2 , C_1]} \rangle \langle q^{[C_2 , C_1]} \vert
\label{Hlinkproj}
\end{eqnarray}
corresponds to the unnormalized local projector $\vert q^{[C_2 , C_1]} \rangle \langle q^{[C_2 , C_1]} \vert $
involving the following ket $ \vert q^{[C_2 , C_1]} \rangle$ associated to the oriented link $[C_2 , C_1] $ 
\begin{eqnarray}
\vert q^{[C_2 , C_1]} \rangle  = \sqrt{ D(C_2 , C_1)  }
 \left[  \sqrt{\frac{\psi_0 (C_1)}{\psi_0 (C_2)} } \vert C_2  \rangle  - \sqrt{\frac{\psi_0 (C_2)}{\psi_0 (C_1)} } \vert C_1  \rangle    \right] 
\label{qcpc}
\end{eqnarray}

The ground state $\vert \psi_0 \rangle$
is orthogonal to all the links-kets of Eq. \ref{qcpc}
\begin{eqnarray}
   \langle q^{[C_2 , C_1]} \vert \psi_0 \rangle = 0
\label{psi0orthoomega}
\end{eqnarray}
and is thus annihilated by all the links-operators $H^{[C_2 , C_1]}  $
\begin{eqnarray}
   H^{[C_2 , C_1]} \vert \psi_0 \rangle = 0
\label{Hlinkannihilation}
\end{eqnarray}
If there are ${\cal N}$ configurations $C$ and ${\cal M}$ oriented links $[C_2 , C_1]$ between them,
where ${\cal M} $ is usually bigger than ${\cal N} $ except on the one-dimensional lattice,
this means that the ground-state $\vert \psi_0 \rangle $ involving
 the ${\cal N} $ components $\psi_0(C) $ of Eq. \ref{psi0}
is the simultaneous ground-state of the $ {\cal M}$ links-operators $H^{[C_2 , C_1]}  $ of Eq. \ref{Hlinkproj} : 
in the literature concerning quantum lattice models, such Hamiltonians are called frustration-free
  (see \cite{frustration_free1,frustration_free2,frustration_free3,frustration_free4,frustration_free5,frustration_free6}
  and references therein), while within the ensemble of these frustration-free Hamiltonians,
   the sub-ensemble of the Hamiltonians related to reversible Markov processes
via similarity transformations
   are often called Rokhsar-Kivelson Hamiltonians for historical reasons related to the field of quantum dimer models (see \cite{RK1,RK2} and references therein).

On the other hand, as recalled in Appendix \ref{app_diffusion},
the quantum Hamiltonians related to reversible diffusion processes via similarity transformations
are second-order differential operators in continuous space that can be factorized into the supersymmetric form
$H =  \vec Q^{\dagger} . \vec Q$ of Eqs \ref{hsusy} in terms of the first order differential operator $ \vec Q$
and its adjoint $ \vec Q^{\dagger}$ of \ref{qsusy}
(see the review \cite{review_susyquantum} on supersymmetric quantum mechanics in continuous space).
In order to make more straighforward the link with this supersymmetric factorization for continuous-space models,
it is thus useful to rewrite the lattice quantum Hamiltonian $H$
 of Eq. \ref{Hsum} of size ${\cal N} \times {\cal N}$ as
 \begin{eqnarray}
H = {\bold Q}^{\dagger} {\bold Q}
\label{Hsusyrectangular}
\end{eqnarray}
where ${\bold Q}$ is the rectangular matrix of size ${\cal M} \times {\cal N}$ whose matrix elements between oriented links $\langle \langle _{C_1}^{C_2} \vert $ and configurations $\vert C \rangle $ are directly related to the links-kets of Eq. \ref{qcpc}
 \begin{eqnarray}
\langle \langle _{C_1}^{C_2} \vert {\bold Q} \vert C \rangle \equiv \langle q^{[C_2 , C_1]} \vert C \rangle
&&  =   \sqrt{ D(C_1,C_2)  } \bigg( \sqrt{\frac{\psi_0 (C_1)}{\psi_0 (C_2)} } \delta_{C_2,C}  -\sqrt{\frac{\psi_0 (C_2)}{\psi_0 (C_1)} }  \delta_{C_1,C}  \bigg) 
\nonumber \\
&& =   \sqrt{ w(C_1,C_2) } \delta_{C_2,C}  -\sqrt{ w(C_2,C_1) }  \delta_{C_1,C} 
\label{Qsusyrectangular}
\end{eqnarray}
The last expression in terms of the detailed-balance transition rates $w(.,.)$ of Eq. \ref{detailedratio}
shows that the rectangular matrix ${\bold Q}$ is a very simple deformation of the incidence matrix ${\bold I}(.,.)$
that is recovered when the transition rates are all unity
  \begin{eqnarray}
\langle \langle _{C_1}^{C_2} \vert {\bold I} \vert C \rangle \equiv    \delta_{C_2,C}  - \delta_{C_1,C} 
\label{Iincidence}
\end{eqnarray}
The application of ${\bold Q}$ to the ground state $  \vert \psi_0 \rangle $
gives zero for any oriented link $\langle \langle _{C_1}^{C_2} \vert $ in consistency with Eq. \ref{psi0orthoomega}
 \begin{eqnarray}
\langle \langle _{C_1}^{C_2} \vert {\bold Q} \vert \psi_0 \rangle 
&& = \sum_C \langle \langle _{C_1}^{C_2} \vert {\bold Q} \vert C \rangle \langle C \vert \psi_0 \rangle
=  \sum_C \sqrt{ D(C_1,C_2)  } \bigg( \sqrt{\frac{\psi_0 (C_1)}{\psi_0 (C_2)} } \delta_{C_2,C}  -\sqrt{\frac{\psi_0 (C_2)}{\psi_0 (C_1)} }  \delta_{C_1,C}  \bigg)  \psi_0(C) 
\nonumber \\
&& =  \sqrt{ D(C_1,C_2)  } \bigg( \sqrt{\frac{\psi_0 (C_1)}{\psi_0 (C_2)} } \psi_0(C_2)  
 -\sqrt{\frac{\psi_0 (C_2)}{\psi_0 (C_1)} }  \psi_0(C_1)   \bigg)  =0
\label{QsusyrectangularGSlink}
\end{eqnarray}
so that the ground state $  \vert \psi_0 \rangle $ that is annihilated by  
the supersymmetric ${\cal N} \times {\cal N}$ Hamiltonian $H = {\bold Q}^{\dagger} {\bold Q} $ is actually annihilated 
by the rectangular matrix ${\bold Q}$ of size ${\cal M} \times {\cal N}$
 \begin{eqnarray}
 {\bold Q} \vert \psi_0 \rangle =0
\label{QsusyrectangularGS}
\end{eqnarray}
which is the analog of Eq. \ref{psi0rannihilate}.


\section{ Large deviations at level 2 for the empirical density } 

\label{sec_level2}

For a Markov-jump trajectory $C(0 \leq t \leq T) $ over the large-time window $t \in [0,T]$,
the empirical density ${\hat p}(C )$ measures the fraction of time spent in each configuration $C$
\begin{eqnarray}
{\hat p}(C ) \equiv \frac{1}{T} \int_0^T dt \delta_{C(t) ,C}
\label{rho1empi}
\end{eqnarray}
so that it satisfies the normalization 
\begin{eqnarray}
\sum_{C} {\hat p}(C ) =1
\label{rho1spinnorma}
\end{eqnarray}
In this section, we recall how the fluctuations of the empirical density ${\hat p}(C )$
around the steady state $p_*(C)$ can be analyzed for large $T$ 
via the rate function $I^{[2]}[{\hat p}(.)   ]  $ that governs the large deviation form
of the distribution ${\cal P}^{[2]}_T[ {\hat p}(.)] $ of ${\hat p}(. )$
\begin{eqnarray}
P^{[2]}_{T}[{\hat p}(.)  ]  \oppropto_{T \to +\infty} 
 \delta \left( \sum_C {\hat p}(C) - 1 \right) 
 e^{- T I^{[2]}[{\hat p}(.)   ] }
\label{level2master}
\end{eqnarray}
We first describe the case of an arbitrary generator $w$ 
in order to stress the simplifications in the presence of detailed balance. 


\subsection{  Reminder on the rate function $I^{[2]}[{\hat p}(.)   ] $ at level 2
for an arbitrary generator $w(.,.)$}

\subsubsection{ Generating function $Z_T^{[\omega(.)]} $ of the empirical density ${\hat p}(.)$ for large $T$  }  

Instead of characterizing the statistics 
of the empirical density ${\hat p}(.)$ 
via its distribution $P^{[2]}_T \left[ {\hat p}(.)  \right] $ of Eq. \ref{level2master},
one can consider its generating function $Z_T^{[\omega(.)]} $ 
where the function $\omega(.) $ is conjugated to the empirical density ${\hat p}(.)$
\begin{eqnarray}
&& Z_T^{[\omega(.)]}  \equiv \int d{\hat p}(.) P^{[2]}_T \left[ {\hat p}(.)  \right]
e^{ \displaystyle T 
 \sum_{C  } {\hat p}(C)  \omega( C)   }
\nonumber \\
&&  \opsimeq_{T \to + \infty} 
\int d{\hat p}(.)   \delta \left(  \sum_{C} {\hat p}(C )-1\right)
e^{ \displaystyle T \left[ 
 \sum_{C  } {\hat p}(C)  \omega( C)    -  I^{[2]}  [ {\hat p}(.)]  \right]  }
 \opsimeq_{T \to + \infty}  
 e^{ \displaystyle T G[\omega(.)] }
\label{geneddomin0saddle}
\end{eqnarray}
On the last line, 
the saddle-point method for large $T$ shows that $G[\omega(.)] $
corresponds to the optimal value of the function in the exponential over the empirical density ${\hat p}(.)$ satisfying the normalization constraint
\begin{eqnarray}
G[\omega(.)]
= \max_{\substack{{\hat p}(.)  \\ \text{satisfying $\sum_{C} {\hat p}(C )=1$} }}
 \left[ 
 \sum_{C  }  {\hat p}(C)  \omega( C)    -  I^{[2]}  [ {\hat p}(.)]  
\right] 
\label{legendrereci}
\end{eqnarray}

\subsubsection{ $G[\omega(.)] $ as the highest eigenvalues of the appropriate $\omega(.)$-deformations of the generator $w(.,.)$ }  

It is useful consider the generalization of the generating function of Eq. \ref{geneddomin0saddle}
where one adds the information on the initial configuration $C(0)=C_0$ and on the final configuration $C(T)=C_T$ 
\begin{eqnarray}
Z_T^{[\omega(.)]}(C_T \vert C_0) 
&&  \equiv \int d{\hat p}(.) \delta_{C(T),C_T} \delta_{C(0),C_0}  P^{[2]}_T \left[ {\hat p}(.)  \right] e^{ \displaystyle T 
 \sum_{ C } {\hat p}(C)  \omega( C)   } 
\nonumber \\
&& = \int_{C(0)=C_0}^{C(T)=C_T}  {\cal D} C(.) \ {\cal P}^{Traj}[C(.) ]   e^{ \displaystyle  \int_0^T dt  \omega( C(t))  } 
\label{geneddef}
\end{eqnarray}
in order to obtain the last expression corresponding to the Markov-jump analog of a Feynman path-integral over the trajectories $C(0 \leq t \leq T) $ starting at $C(0)=C_0$ and ending at $C(T)=C$ 
with their probabilities ${\cal P}^{Traj}[C(0 \leq t \leq T)] $ that read in terms of the generator $w(.,.)$
\begin{eqnarray}
{\cal P}^{Traj}[C(0 \leq t \leq T)]   
=   e^{ \displaystyle    \sum_{t \in [0,T]: C(t^+) \ne C(t) } \ln (w(C(t^+),C(t))  ) +  \int_0^T dt w(C(t),C(t))   }
\label{pwtrajjump}
\end{eqnarray}
As a consequence, the generating function of Eq. \ref{geneddef} becomes
\begin{eqnarray}
Z_T^{[\omega(.)]}(C_T \vert C_0) 
&& = \int_{C(0)=C_0}^{C(T)=C_T}  {\cal D} C(.) \ 
e^{ \displaystyle    \sum_{t \in [0,T]: C(t^+) \ne C(t) } \ln (w(C(t^+),C(t))  ) +  \int_0^T dt \left[ w(C(t),C(t)) +\omega( C(t) \right]  }
\nonumber \\
&& 
 =  \langle C_T \vert e^{T w^{[\omega(.)]} } \vert C_0\rangle
\label{geneddeformed}
\end{eqnarray}
that corresponds to the propagator $\langle C_T \vert e^{T w^{[\omega(.)]} } \vert C_0\rangle $
 associated to the following deformed-matrix
$w^{[\omega(.)]} $ with respect to the initial Markov generator $w(.,.)$ of Eq. \ref{master} 
\begin{eqnarray}
w^{[\omega(.)]}(C,C') && \equiv w(C ,C')  + \omega(C)\delta_{C,C'}
\label{markovmatrixdeformed}
\end{eqnarray}
This is the analog of the famous Feynman-Kac formula for diffusion processes.

The Perron-Frobenius theorem
yields that the leading behavior of the generating function of Eq. \ref{geneddeformed} for large $T$
\begin{eqnarray}
Z_T^{[\omega(.)]}(C_T \vert C_0) 
  \opsimeq_{T \to + \infty} 
 e^{T G[\omega(.)] } \ \ r^{[\omega(.)]}(C_T) \ \  l^{[\omega(.)]}(C_0)
\label{geneddomin0}
\end{eqnarray}
is governed by the highest eigenvalue $G[\omega(.)]$ of the deformed-matrix
$w^{[\omega(.)]} $ :  the positive left eigenvector $l^{[\omega(.)]}(C)  \geq 0$ 
corresponds to the deformation of the trivial eigenvector $l_0(C)=1$ of Eq. \ref{markovleftj}
\begin{eqnarray}
G[\omega(.)]   l^{[\omega(.)]}(  C')   
&& =\sum_C  l^{[\omega(.)]}(  C) w^{[\omega(.) ]}(C, C')
\nonumber \\
&& = l^{[\omega(.)]}(  C') \left[ \omega(C')+w(C',C')\right]
+ \sum_{C \ne C'}  l^{[\omega(.)]}(  C) 
w(C , C') 
 \label{eigenleftwdef}
\end{eqnarray}
while the positive right 
eigenvector $ r^{[\omega(.)]}(C)   \geq 0 $ 
corresponds to the deformation of the right eigenvector $r_0(C)=p_*(C)$ of Eq. \ref{markovrightj}
\begin{eqnarray}
G[\omega(.)]    r^{[\omega(.)]}(  C) 
&&  = \sum_{C'} w^{[\omega(.)]}(C, C')r^{[\omega(.)]}(  C') 
\nonumber \\
&& =   \left[  \omega(C)+w(C,C)\right] r^{[\omega(.)]}(  C) 
+  \sum_{C' \ne C} w(C , C')  r^{[\omega(.)]}(  C') 
\label{eigenrightwdef}
\end{eqnarray}
with the normalization
\begin{eqnarray}
1 = \langle l^{[\omega(.)]}  \vert r^{[\omega(.)]} \rangle 
= \sum_C l^{[\omega(.)]}(C)  r^{[\omega(.)]}(C) 
 \label{Wknorma}
\end{eqnarray}


\subsubsection{  Inverse Functional Legendre transform to obtain the rate function $ I^{[2]}  [ {\hat p}(.)]  $ 
from the eigenvalues $G[\omega(.)]  $  }

The reciprocal Legendre transformation of Eq. \ref{legendrereci} 
\begin{eqnarray}
I^{[2]}  [ {\hat p}(.)]
= \max_{\omega(.)  }
 \left[ 
 \sum_{C  } {\hat p}(C)  \omega( C)    -  G[\omega(.)]
\right] 
\label{legendre}
\end{eqnarray}
involves the optimization with respect to $\omega(C)$ 
\begin{eqnarray}
0=\frac{ \partial }{\partial \omega(C)  }
 \left[ 
 \sum_{C'  } {\hat p}(C')  \omega( C')    -  G[\omega(.)]
\right] 
= {\hat p}(C) - \frac{ \partial G[\omega(.)]}{\partial \omega(C)  }
\label{legendrederi}
\end{eqnarray}
where one needs to evaluate the
derivative with respect to $\omega(C)$ of the eigenvalue
\begin{eqnarray}
 G[\omega(.)] = \langle l^{[\omega(.)]} \vert w^{[\omega(.) ]} \vert r^{[\omega(.)]} \rangle
 = \sum_C \sum_{C'}  l^{[\omega(.)]} (C)  \left[ w(C , C')  + \omega(C)\delta_{C,C'} \right]   r^{[\omega(.)]} (C')
\label{Geigenomega}
\end{eqnarray}
As a consequence of the normalization of Eq \ref{Wknorma} valid for any $\omega(.)$, 
only the derivative of the deformed-matrix $w^{[\omega(.) ]} $ survives 
\begin{eqnarray}
\frac{ \partial G[\omega(.)]}{\partial \omega(C)  } && = 
\langle l^{[\omega(.)]} \vert  \left( \frac{ \partial w^{[\omega(.) ]} }{\partial \omega(C)  } \right)\vert r^{[\omega(.)]} \rangle
 = 
 \sum_{C'}  l^{[\omega(.)]} (C)  \delta_{C,C'}  r^{[\omega(.)]} (C')
 =    l^{[\omega(.)]} (C)    r^{[\omega(.)]} (C)
\label{Geigenomegaderi}
\end{eqnarray}
So the optimization of Eq. \ref{legendrederi} yields that 
the empirical density ${\hat p}(C)  $ corresponds to the product of the left and right eigenvectors
\begin{eqnarray}
{\hat p}(C) = \frac{ \partial G[\omega(.)]}{\partial \omega(C)  } =  l^{[\omega(.)]} (C)    r^{[\omega(.)]} (C)
\label{legendrederieigen}
\end{eqnarray}

One can use the eigenvalue Eq. \ref{eigenleftwdef} for the left eigenvector $l^{[\omega(.)]}(  C) $ 
to replace
\begin{eqnarray}
\omega(C) - G[\omega(.)]  
 =  -  \sum_{ C'}  l^{[\omega(.)]}(  C') w(C' , C) \frac{1}{ l^{[\omega(.)]}(  C)}
 \label{eigenleftwdefG}
\end{eqnarray}
into Eq. \ref{legendre}
to rewrite the rate function at level 2 as
\begin{eqnarray}
I^{[2]}  [ {\hat p}(.)]
= \max_{\omega(.)  }
 \left[  \sum_{C  }  \bigg( \omega( C)    -  G[\omega(.)] \bigg) {\hat p}(C) \right] 
= \max_{\omega(.)  }
 \left[ - \sum_{C  }   \sum_{ C'}  l^{[\omega(.)]}(  C') w(C' , C) \frac{1}{ l^{[\omega(.)]}(  C)}  {\hat p}(C) \right] 
\label{legendreleft}
\end{eqnarray}
 One can use Eq. \ref{legendrederieigen} to obtain the matrix element 
 of the generator $w$ between the left and right eigenvectors
\begin{eqnarray}
I^{[2]}  [ {\hat p}(.)]
&& = \max_{\omega(.) :  {\hat p}(.)= l^{[\omega(.)]} (.)    r^{[\omega(.)]} (.) }
 \left[ - \sum_{C  }   \sum_{ C'}  l^{[\omega(.)]}(  C') w(C' , C) r^{[\omega(.)]}(  C) \right] 
\nonumber \\
&&  =\max_{\omega(.)  :  {\hat p}(.)= l^{[\omega(.)]} (.)    r^{[\omega(.)]} (.)}
 \left[ - \langle l^{[\omega(.)]} \vert w \vert r^{[\omega(.)]} \rangle \right] 
\label{legendreleftright}
\end{eqnarray}
However for arbitrary irreversible generators,  
one prefers to keep Eq. \ref{legendreleft}, where
the only dependence with respect to $\omega(.)$
 is contained is the single positive left eigenvector $l^{[\omega(.)]}(  .) $,
 so that the optimization of Eq. \ref{legendreleft} is actually only over some positive function $l(.)$
 and one can forget the function $\omega(.)$ to obtain 
 the Donsker-Varadhan optimization-formula \cite{DonskerV}
\begin{eqnarray}
I^{[2]}  [ {\hat p}(.)]
= \max_{l(.) >0 }
 \left[ - \sum_{C'  }   \sum_{ C}  l(  C') w(C' , C) \frac{1}{ l(  C)}  {\hat p}(C) \right] 
\label{legendreleftl}
\end{eqnarray}
which is still not explicit as a function of the empirical density ${\hat p}(.) $ for an arbitrary irreversible generator $w$,
as a consequence of the remaining optimization  
with respect to the positive function $l(.)$.


\subsubsection{ Canonical conditioning with respect to the empirical density ${\hat p}(.)$ }

Since the generating function $Z_T^{[\omega(.)]}(C \vert C_0)$ of Eq. \ref{geneddomin0}
will grow or decay exponentially in time whenever the eigenvalue $G[\omega(.)] $
is non-vanishing,
it is useful to construct the conditioned propagator (see the two very detailed papers \cite{chetrite_conditioned,chetrite_optimal} and references therein)
\begin{eqnarray}
P_T^{Cond[\omega(.)]}(C \vert C_0) && \equiv e^{- T G[\omega(.)] } 
\frac{ l^{[\omega(.)]}(C)}{ l^{[\omega(.)]}(C_0)} Z_T^{[\omega(.)]}(C \vert C_0) 
\nonumber \\
&&  \opsimeq_{T \to + \infty} 
 l^{[\omega(.)]}(C) \ \  r^{[\omega(.)]}(C) \equiv
p^{Cond[\omega(.)]}_* (C)
\label{conditionedpropagator}
\end{eqnarray}
that converges, independently of the initial condition $C_0$,
 towards the conditioned steady state $p^{Cond[\omega(.)]}_* (C) $
given by the product of the left eigenvector $l^{[\omega(.)]}(C) $ of Eqs  \ref{eigenleftwdef}
and the right eigenvector $r^{[\omega(.)]}(C) $ of Eq \ref{eigenrightwdef},
i.e. one recognizes the empirical steady state ${\hat p }(C) $ of Eq. \ref{legendrederieigen}
\begin{eqnarray}
p^{Cond[\omega(.)]}_* (C) =  l^{[\omega(.)]}(C) \ \  r^{[\omega(.)]}(C) = {\hat p }(C)
\label{CondSteady}
\end{eqnarray}

The corresponding conditioned Markov generator $w^{Cond[\omega(.)]}(.,.) $ 
can be constructed from the deformed matrix $w^{[\omega(.)]}(C , C') $ of Eq. \ref{markovmatrixdeformed} as follows:

(i) The off-diagonal element $w^{Cond[\omega(.)]}(C,C')$ for $C \ne C'$
\begin{eqnarray}
{\rm for } \ \ C \ne C' : \ \ w^{Cond[\omega(.)]}(C,C') 
&& =   l^{[\omega(.)]}(C)  w^{[\omega(.)]}(C , C')  \frac{1}{l^{[\omega(.)]}(C')}  
\nonumber \\
&& =   l^{[\omega(.)]}(C)  w(C , C')  \frac{1}{l^{[\omega(.)]}(C')} 
\label{wjumpforwardklargedev}
\end{eqnarray}
only involves the initial off-diagonal element $w(C,C') $ and 
the left eigenvector $l^{[\omega(.)]}(.) $.

(ii) The diagonal element $w^{Cond[\omega(.)]}(C,C) $ is determined in terms of the off-diagonal elements 
$w^{Cond[\omega(.)]}(C',C)  $ with $C' \ne C$ by the conservation of probability analog to Eq. \ref{wdiag}
and can be rewritten using the left eigenvalue Eq. \ref{eigenleftwdef} 
\begin{eqnarray}
w^{Cond[\omega(.)]}(C,C) && = - \sum_{C' \ne C} w^{Cond[\omega(.)]}(C',C) 
 = -  \left[ \sum_{C' \ne C}  l^{[\omega(.)]}(C')  w(C' , C)  \right]
 \frac{1}{l^{[\omega(.)]}(C)}
 \nonumber \\ &&
 = -  \left[ G[\omega(.)]   l^{[\omega(.)]}(  C) -  l^{[\omega(.)]}(C)  w^{[\omega(.)]}(C , C)  \right]   \frac{1}{l^{[\omega(.)]}(C)}
\nonumber \\
&& = w^{[\omega(.)]}(C , C) - G[\omega(.)]
\nonumber \\
&& = w(C,C) + \omega(C)- G[\omega(.)]
\label{wjumpforwardklargedevdiag}
\end{eqnarray}
in terms
the initial diagonal element $w(C,C) $, the function $\omega(C) $ and the eigenvalue $G[\omega(.)] $.

In conclusion, for an arbitrary irreversible generator $w$, 
one needs to be able to compute the eigenvalue $G[\omega(.)] $ and the left eigenvector $l^{[\omega(.)]}(.) $
to compute the conditioned generator $w^{Cond[\omega(.)]}(.,.) $ via Eqs \ref{wjumpforwardklargedev} and \ref{wjumpforwardklargedevdiag}.


\subsection{ Simplification for the rate function $I^{[2]}  [ {\hat p}(.)] $ and for the conditioning when the generator 
is reversible  }

In this subsection, we describe how the general framework recalled in the previous subsection
for an arbitrary generator $w$ leads to explicit results in the presence of detailed-balance. 

\subsubsection{ Explicit rate function $I^{[2]}  [ {\hat p}(.)] $ of the empirical density ${\hat p}(.) $ 
when the generator $w(.,.)$
is reversible }

Since the deformed-matrix $w^{[\omega(.)]}(C,C') $ of Eq. \ref{markovmatrixdeformed}
only involves a diagonal deformation,
one can make the same similarity transformation as in Eq. \ref{Hsimilarity}
for the undeformed reversible generator $w$
to obtain the deformed symmetric quantum Hamiltonian
\begin{eqnarray}
H^{[\omega(.)]}(C, C') && \equiv  - \frac{1}{ \sqrt{p_* (C) }} w^{[\omega(.)]}(C,C') \sqrt{p_* (C') }
=  - \frac{1}{ \sqrt{p_* (C) }} \left[ w(C ,C')  + \omega(C)\delta_{C,C'}\right] \sqrt{p_* (C') }
\nonumber \\
&& = H(C , C')  - \omega(C)\delta_{C,C'} = H^{[\omega(.)]}(C', C) 
\label{Hdeformed}
\end{eqnarray}
As a consequence, the left and right positive eigenvectors of Eqs \ref{eigenleftwdef} \ref{eigenrightwdef}
for the deformed-matrix $w^{[\omega(.)]}(C,C') $
can be rewritten as
\begin{eqnarray}
r^{[\omega(.)]}(C) && = \sqrt{p_* (C) }  \psi^{[\omega(.)]} ( C )
\nonumber \\
l^{[\omega(.)]}(C) && = \frac{ \psi^{[\omega(.)]} ( C ) }{\sqrt{p_* (C) } }
\label{rldeformedpsi}
\end{eqnarray}
in terms of the positive ground-state $\psi^{[\omega(.)]} (.)$ 
of the deformed Hamiltonian of Eq. \ref{Hdeformed}
associated to the eigenvalue $[- G[\omega(.)]] $ 
\begin{eqnarray}
 H^{[\omega(.)]} \vert \psi^{[\omega(.)]} \rangle && =  - G[\omega(.)] \vert \psi^{[\omega(.)]} \rangle
\label{eigenpsi}
\end{eqnarray}
The empirical density of Eq. \ref{legendrederieigen} given by the product of the left and right eigenvectors of Eq. \ref{rldeformedpsi}
reduces to the square of the deformed ground-state $\psi^{[\omega(.)]} (.) $.
\begin{eqnarray}
{\hat p}(C) =  l^{[\omega(.)]} (C)    r^{[\omega(.)]} (C) = \left[  \psi^{[\omega(.)]} ( C )\right]^2
\label{empicondi}
\end{eqnarray}
while the matrix element appearing in Eq. \ref{legendreleftright}
can be rewritten in terms of the undeformed quantum Hamiltonian $H$ and 
the deformed ground-state $\psi^{[\omega(.)]} ( C ) $ as
\begin{eqnarray}
 \langle l^{[\omega(.)]} \vert w \vert r^{[\omega(.)]} \rangle
&& = \sum_{C  }   \sum_{ C'}  l^{[\omega(.)]}(  C') w(C' , C) r^{[\omega(.)]}(  C)
 = \sum_{C  }   \sum_{ C'}  \frac{ \psi^{[\omega(.)]} ( C' ) }{\sqrt{p_* (C') } } w(C' , C) \sqrt{p_* (C) }  \psi^{[\omega(.)]} ( C )
\nonumber \\
&&  = - \sum_{C  }   \sum_{ C'}  \psi^{[\omega(.)]} ( C' ) H(C' , C)   \psi^{[\omega(.)]} ( C )
 = -  \langle \psi^{[\omega(.)]} \vert H \vert \psi^{[\omega(.)]} \rangle
\label{legendreleftrightmatrix}
\end{eqnarray}
As a consequence, the rate function of Eq. \ref{legendreleftright} 
\begin{eqnarray}
I^{[2]}  [ {\hat p}(.)]
= \max_{\omega(.) :  {\hat p}(.)= [\psi^{[\omega(.)]} (.) ]^2 }
 \left[ \langle \psi^{[\omega(.)]} \vert H \vert \psi^{[\omega(.)]} \rangle \right] 
 = \langle \sqrt{ {\hat p}} \vert H \vert \sqrt{ {\hat p}} \rangle
\label{legendreleftrightexplicit}
\end{eqnarray}
reduces to the explicit matrix element $\langle \sqrt{ {\hat p}} \vert H \vert \sqrt{ {\hat p}} \rangle $
that involves only the quantum supersymmetric Hamiltonian $H$ and the square-root ket $\vert \sqrt{ {\hat p}} \rangle $.
So in contrast to the Donsker-Varadhan optimization-formula of Eq. \ref{legendreleftl} for irreversible processes
that is not explicit, the Donsker-Varadhan formula for the rate function at level 2 is explicit
for reversible processes, and only involves the supersymmetric Hamiltonian $H$.

Let us now stress some important consequences of this formula:

(i) If one replaces $H$ by its spectral decomposition of Eq. \ref{Hspectral},
the rate function of Eq. \ref{legendreleftrightexplicit}
becomes
\begin{eqnarray}
I^{[2]}  [ {\hat p}(.)]   = \sum_{n=1}^{{\cal N}-1} E_n \langle \sqrt{ {\hat p}}\vert \psi_n \rangle  \langle \psi_n \vert\sqrt{ {\hat p}} \rangle
 = \sum_{n=1}^{{\cal N}-1} E_n \left(  \langle \psi_n \vert\sqrt{ {\hat p}} \rangle \right)^2
\label{I2spectral}
\end{eqnarray}
that involves the $({\cal N}-1)$ excited energies $E_n>0$ and the 
weights $  \left(  \langle \psi_n \vert\sqrt{ {\hat p}} \rangle \right)^2 $ 
of the square-root ket $\vert\sqrt{ {\hat p}} \rangle $ on the corresponding excited eigenstates $\psi_n(C)$
with the normalization from Eq. \ref{orthopsi}
\begin{eqnarray}
1   = \langle \sqrt{ {\hat p}} \vert \sqrt{ {\hat p}} \rangle 
= \langle \sqrt{ {\hat p}} \vert \bigg( \sum_{n=0}^{{\cal N}-1}  \vert \psi_n \rangle  \langle \psi_n \vert \bigg) \vert \sqrt{ {\hat p}} \rangle 
 = \left(  \langle \psi_0 \vert\sqrt{ {\hat p}} \rangle \right)^2 + \sum_{n=1}^{{\cal N}-1}  \left(  \langle \psi_n \vert\sqrt{ {\hat p}} \rangle \right)^2
\label{normaweightswn}
\end{eqnarray}

(ii) If one replaces the hamiltonian $H$ by its supersymmetric form of Eqs \ref{Hsum} \ref{Hlinkproj},
the rate function of Eq. \ref{legendreleftrightexplicit}
becomes a sum over the links of the positive contributions $\left( \langle q^{[C' , C]} \vert \sqrt{ {\hat p}} \rangle \right)^2 $ 
corresponding to the square of the amplitude of $\vert \sqrt{ {\hat p}} \rangle  $ on the link-kets $\vert q^{[C' , C]} \rangle $
\begin{eqnarray}
I^{[2]}  [ {\hat p}(.)]
&& = \sum_{C} \sum_{C'>C} \langle \sqrt{ {\hat p}} \vert q^{[C' , C]} \rangle \langle q^{[C' , C]} \vert \sqrt{ {\hat p}} \rangle
 =  \sum_{C} \sum_{C'>C} \left( \langle q^{[C' , C]} \vert \sqrt{ {\hat p}} \rangle \right)^2  
 \nonumber \\
 && =  \sum_{C} \sum_{C'>C}  D(C',C)  
\left[ \sqrt{\frac{\psi_0 (C)}{\psi_0 (C')} } \sqrt{ {\hat p}(C') }  - \sqrt{\frac{\psi_0 (C')}{\psi_0 (C)} } \sqrt{ {\hat p} (C)}    \right]^2
 \nonumber \\
 && =  \sum_{C} \sum_{C'>C}  D(C',C)  \sqrt{ p_* (C) p_* (C') }
\left[ \sqrt{\frac{ {\hat p}(C')}{p_* (C')} }   - \sqrt{\frac{ {\hat p}(C)}{p_* (C)} }    \right]^2
\label{I2assumofsquares}
\end{eqnarray}
 where the last line was rewritten in terms of the steady state $p_*(C)=\psi_0^2(C)$ to show more clearly the cost
 of an empirical density $ {\hat p} (.)$ different from the steady state $p_*(.)$,
 and to make the link with the formula of Eq. \ref{rate2diff} concerning diffusion processes.

(iii) Another useful rewriting of the rate function given by the first line of Eq. \ref{I2assumofsquares}
\begin{eqnarray}
I^{[2]}  [ {\hat p}(.)]
 = \sum_{C} \sum_{C'>C} \langle q^{[C' , C]} \vert \sqrt{ {\hat p}} \rangle \langle \sqrt{ {\hat p}} \vert q^{[C' , C]} \rangle 
 \equiv  \sum_{C} \sum_{C'>C} \langle q^{[C' , C]} \vert {\hat \rho} \vert q^{[C' , C]} \rangle 
\label{I2densitymatrix}
\end{eqnarray}
involves the empirical quantum density matrix $ {\hat \rho} $ associated to the square-root state $\vert \sqrt{ {\hat p}} \rangle $
\begin{eqnarray}
 {\hat \rho} \equiv \vert \sqrt{ {\hat p}} \rangle  \langle \sqrt{ {\hat p}} \vert 
\label{rhotot}
\end{eqnarray}
If there are ${\cal N}$ configurations $C$, the quantum density matrix ${\hat \rho} $ is of size ${\cal N} \times {\cal N}$,
where the ${\cal N}$ diagonal elements simply correspond to the empirical density ${\hat p}(C)   $ of Eq. \ref{rho1empi}
\begin{eqnarray}
\langle C \vert {\hat \rho} \vert C \rangle = \langle C\vert \sqrt{ {\hat p}} \rangle \langle \sqrt{ {\hat p}} \vert C \rangle
= {\hat p}(C) = \frac{1}{T} \int_0^T dt \delta_{C(t) ,C} 
\label{rhototdiag}
\end{eqnarray}
with the corresponding normalization concerning the trace of ${\hat \rho} $ 
\begin{eqnarray}
1=\sum_C   {\hat p}(C) = \sum_C \langle C \vert {\hat \rho} \vert C \rangle = \text{Tr} ({\hat \rho}  ) 
\label{rhototdiagnorma}
\end{eqnarray}
while the symmetric off-diagonal elements $C' \ne C$ read
\begin{eqnarray}
\langle C' \vert {\hat \rho} \vert C \rangle  = \langle C \vert {\hat \rho} \vert C' \rangle
=  \sqrt{ {\hat p}(C') {\hat p}(C) }  = 
 \frac{1}{T} \sqrt{ \left( \int_0^T dt' \delta_{C(t') ,C'} \right) \left( \int_0^T dt \delta_{C(t) ,C} \right) } 
\label{rhototoff}
\end{eqnarray}
The expression of Eq. \ref{I2densitymatrix} means that the rate function $I^{[2]}  [ {\hat p}(.)] $
only involves:

(a) the ${\cal N}$ diagonal elements of Eq. \ref{rhototdiag} 

(b) the off-diagonal elements $C' \ne C$ of Eq. \ref{rhototoff}
that are related by non-vanishing transition rates $D(C,C')>0 $.

This formulation in terms of the density matrix ${\hat \rho} = \vert \sqrt{ {\hat p}} \rangle  \langle \sqrt{ {\hat p}} \vert $
of Eq. \ref{rhotot} is useful for manybody models with local transition rates,
as will be discussed in section \ref{sec_spins} on examples of pure or random spin chains.


\subsubsection{ Explicit conditioned generator of a reversible generator $w$ with respect to the empirical density ${\hat p}(.)$ } 

\label{subsec_conditioning}

When the generator $w$ satisfies detailed-balance,
 one can plug the parametrization of Eq. \ref{detailedratio}
 and the left eigenvector of Eq. \ref{rldeformedpsi}
into Eq. \ref{wjumpforwardklargedev}
to obtain that the off-diagonal elements of the conditioned generator
\begin{eqnarray}
{\rm for } \ \ C \ne C' : \ \ w^{Cond[\omega(.)]}(C,C') 
&& =  l^{[\omega(.)]}(C)  w(C , C')  \frac{1}{l^{[\omega(.)]}(C)} 
 =   \frac{ \psi^{[\omega(.)]} ( C ) }{\sqrt{p_* (C) } } \times
  \sqrt{p_*(C)} D(C , C') \frac{1}{ \sqrt{p_*(C') }} 
 \times \frac{1}{  \frac{ \psi^{[\omega(.)]} ( C' ) }{\sqrt{p_* (C') } }} 
\nonumber \\
&& = \psi^{[\omega(.)]} ( C ) D(C , C') \frac{1}{ \psi^{[\omega(.)]} ( C' ) } 
 \equiv \sqrt{{\hat p}(C)} D(C , C')  \frac{1}{ \sqrt{{\hat p} (C')}} 
\label{wjumpforwardklargedevrev}
\end{eqnarray}
satisfy detailed-balance with respect to the conditioned steady state given by the empirical density
\begin{eqnarray}
{\hat p} (C) =p^{Cond[\omega(C)]} = \left[ \psi^{[\omega(.)]} ( C )\right]^2
\label{conditionedsteadyrev}
\end{eqnarray}
The diagonal elements of the conditioned generator can be then obtained from the off-diagonal elements of Eq. \ref{wjumpforwardklargedevrev}
\begin{eqnarray}
w^{Cond[\omega(.)]}(C',C')  = - \sum_{C \ne C'} w^{Cond[\omega(.)]}(C,C') 
= - \sum_{C \ne C'} \sqrt{{\hat p}(C)} D(C , C')  \frac{1}{ \sqrt{{\hat p} (C')}} 
\label{wjumpforwardklargedevdiagrev}
\end{eqnarray}

In summary, when the generator $w$ is reversible, 
even if one cannot solve explicitly the eigenvalue equation for the deformed-matrix 
$w^{[\omega(.)]}(C',C')  $ of Eq. \ref{markovmatrixdeformed},
one can nevertheless write the explicit expression of Eqs \ref{wjumpforwardklargedevrev} \ref{wjumpforwardklargedevdiagrev}
for the conditioned generator that will produce the given empirical density ${\hat p}(.) $ : 
it is thus clearer to forget the function $\omega(.)$ and to write that the 
generator $w^{Cond[\hat p(.)]} $ conditioned to produce the given empirical density ${\hat p}(.) $ reads
\begin{eqnarray}
w^{Cond[\hat p(.)]}(C,C') 
&& =  D(C , C') \sqrt{\frac{ {\hat p}(C)}{ {\hat p} (C')} }  \ \ \ {\rm for } \ \ C \ne C'
 \nonumber \\
w^{Cond[\hat p(.)]}(C',C') && = - \sum_{C \ne C'} w^{Cond[\hat p(.)]}(C,C') 
= - \sum_{C \ne C'}  D(C , C') \sqrt{\frac{ {\hat p}(C)}{ {\hat p} (C')} }
\label{wconditionedp}
\end{eqnarray}
Since this conditioned generator $w^{Cond[\hat p(.)]}(C,C')  $
satisfies detailed-balance with respect to the empirical density ${\hat p}(.) $,
it is useful to write the corresponding supersymmetric quantum Hamiltonian via the 
similarity transformation analog of Eq. \ref{Hsimilarity}
\begin{eqnarray}
 H^{Cond[{\hat p}(.)]}(C,C') 
 = - \frac{1}{ \sqrt{ {\hat p}(C) }}
 w^{Cond[{\hat p} (.)]}(C,C') \sqrt{ {\hat p}(C') }
\label{Hsimilaritycond}
\end{eqnarray}
So the off-diagonal elements are the same as for the initial Hamiltonian $H$ of Eq. \ref{offdiag}
\begin{eqnarray}
H^{Cond[{\hat p}(.)]}(C,C') = - D(C,C')  = H(C',C)  \ \ \ \text{ for } \ \  C' \ne C
\label{offdiagcond}
\end{eqnarray}
and the only changes occur in the diagonal elements 
\begin{eqnarray}
H^{Cond[{\hat p}(.)]}(C',C')  = - w^{Cond[{\hat p}(.)]}(C',C') 
= \sum_{C \ne C'}  D(C , C') \sqrt{\frac{ {\hat p}(C)}{ {\hat p} (C')} }
\label{diagCond}
\end{eqnarray}


\section{Large deviations at levels higher than 2 for Markov jump processes  }

\label{sec_activities}

For a trajectory $C(0 \leq t \leq T)$ over the large-time window $t \in [0,T]$ of the Markov jump process, 
besides the empirical density ${\hat p}(.)$  of Eq. \ref{rho1empi},
it is also interesting to consider the empirical transition rates as measured from the observed jumps between different configurations $C' \ne C$
\begin{eqnarray}
{\hat w}(C',C ) \equiv \frac{ \displaystyle \frac{1}{T} 
\sum_{t \in [0,T] :C(t^+) \ne C(t) }
\delta_{C(t^+) ,C'} \delta_{C(t) ,C} }
{ {\hat p}(C )} 
\label{Qempi}
\end{eqnarray}
In this section, we describe how one can analyze the fluctuations of this empirical dynamics around the true dynamics
for large $T$, first for the case of an arbitrary generator $w$ and then in the presence of detailed balance.


\subsection{Reminder on the explicit large deviations at level 2.5 for any Markov jump process    }

\subsubsection{Stationarity constraints for the empirical transitions rates $ {\hat w}(.,. )$ or for the empirical currents $ {\hat j}(.,.)$ }

In a Markov jump trajectory $C(0 \leq t \leq T) $, 
the difference between the number of arrivals into a given configuration $C$
and the number of departures from this same configuration $C$ reads in terms of the 
the empirical density ${\hat p}(.)$  of Eq. \ref{rho1empi}
and the empirical transition rates of Eq. \ref{Qempi} 
\begin{eqnarray}
\sum_{t \in [0,T] :C(t^+) \ne C(t) }\delta_{C(t^+) ,C} 
- \sum_{t \in [0,T] :C(t^+) \ne C(t) } \delta_{C(t) ,C} 
= T \left[ \sum_{C'} {\hat w}(C,C') {\hat p}(C' ) - \sum_{C'} {\hat w}(C',C) {\hat p}(C )\right] 
\label{Arrivaldeparture}
\end{eqnarray}
On the other hand, this difference can only differ from zero by the boundary contributions $[ \delta_{C,C(T)} - \delta_{C,C(0)} ] $ of order $O(1)$
involving the initial configuration $C(0)$ and the final configuration $C(T)$,
so that the coefficient of the large time $T$ on the right handside Eq. \ref{Arrivaldeparture} should vanish for any $C$
\begin{eqnarray}
0 = \sum_{C' \ne C} \left[ {\hat w}(C,C' ){\hat p}(C' ) - {\hat w}(C',C ){\hat p}(C )   
 \right] 
\label{statioempi}
\end{eqnarray}
This stationarity constraint means the empirical density ${\hat p}(C) $ should be the steady state of the empirical dynamics governed by the empirical transition rates ${\hat w}(C',C ) $.

Instead of the empirical transition rates ${\hat w}(C',C ) $,
one can introduce the symmetric empirical activities ${\hat a} (C,C') =  {\hat a} (C',C)$ 
and the antisymmetric empirical currents ${\hat j} (C,C')=- {\hat j} (C',C)$
\begin{eqnarray}
{\hat a} (C,C') && \equiv {\hat w}(C,C' ){\hat p}(C' ) + {\hat w}(C',C ){\hat p}(C )   =  {\hat a} (C',C)
\nonumber \\
{\hat j} (C,C') && \equiv {\hat w}(C,C' ){\hat p}(C' ) - {\hat w}(C',C ){\hat p}(C )   = - {\hat j} (C',C)
\label{jempi}
\end{eqnarray}
so that the constitutive constraints of Eq. \ref{statioempi}
reduce to the divergenceless constraint for the empirical currents
\begin{eqnarray}
0 =  \sum_{C' \ne C} {\hat j} (C,C')
\label{statioempij}
\end{eqnarray}


\subsubsection{ Explicit rate function $I^{[2.5]}[ {\hat p}(.); {\hat w}(. , .)]  $ at level 2.5 }

For any Markov jump process,
the joint distribution ${\cal P}^{[2.5]}_T[ {\hat p}(.); {\hat w}(. , .)]  $ of the empirical density ${\hat p}(.)$ 
satisfying the normalization of Eq. \ref{rho1spinnorma}
and of the empirical transition rates ${\hat w}(.,.)$ satisfying the constitutive constraints 
of Eq. \ref{statioempi}
follows the large deviation form for large $T$
\cite{fortelle_thesis,fortelle_jump,maes_canonical,maes_onandbeyond,wynants_thesis,chetrite_formal,BFG1,BFG2,chetrite_HDR,c_ring,c_interactions,c_open,barato_periodic,chetrite_periodic,c_reset,c_inference,c_LargeDevAbsorbing,c_microcanoEnsembles,c_susyboundarydriven,c_Inverse}
\begin{eqnarray}
{\cal P}^{[2.5]}_T[ {\hat p}(.); {\hat w}(. , .)] && 
\opsimeq_{T \to + \infty}  \delta \left(  \sum_{C} {\hat p}(C )-1\right)
\left[ \prod_{C} \delta \left( \sum_{C' \ne C} \left[ {\hat w}(C',C ){\hat p}(C )   -
{\hat w}(C,C' ){\hat p}(C' ) \right]   \right) \right]
  e^{ \displaystyle - T  I^{[2.5]}[ {\hat p}(.); {\hat w}(. , .)]}
  \ \ \ 
\label{proba2.5jumpPw}
\end{eqnarray}
where the explicit rate function at level 2.5
\cite{fortelle_thesis,fortelle_jump,maes_canonical,maes_onandbeyond,wynants_thesis,chetrite_formal,BFG1,BFG2,chetrite_HDR,c_ring,c_interactions,c_open,barato_periodic,chetrite_periodic,c_reset,c_inference,c_LargeDevAbsorbing,c_microcanoEnsembles,c_susyboundarydriven,c_Inverse}
\begin{eqnarray}
I^{[2.5]}[ {\hat p}(.); {\hat w}(. , .)] 
= \sum_{C} {\hat p}(C ) \sum_{C' \ne C } 
\left[  {\hat w}(C' , C )  \ln \left( \frac{ {\hat w}(C', C ) }{w(C' , C ) } \right)
- {\hat w}(C', C ) + w(C' , C )
\right]
\label{rate2.5jumpPw}
\end{eqnarray}
is the relative entropy of the empirical Poisson dynamics with respect to the true Poisson dynamics,
while full entropies can be also be written explicitly (see for instance the detailed discussions in \cite{c_microcanoEnsembles}).

For each link $C'>C$, if one replaces the two empirical transition rates ${\hat w}(C',C ) $ 
and ${\hat w}(C,C' ) $ in terms of the empirical activity ${\hat a} (C,C') $ 
and the empirical current ${\hat j} (C,C')$ of the link given by Eq. \ref{jempi}
\begin{eqnarray}
{\hat w}(C,C' ) && = \frac{{\hat a} (C,C') + {\hat j} (C,C') }{ 2 {\hat p}(C' ) }
\nonumber \\
{\hat w}(C',C ) && = \frac{ {\hat a} (C,C') - {\hat j} (C,C') }{ 2 {\hat p}(C )  }
\label{jempiinverse}
\end{eqnarray}
the large deviations of Eq. \ref{proba2.5jumpPw}
become for the joint distribution ${\cal P}^{[2.5]}_T[ {\hat p}(.); {\hat a}(. , .); {\hat j}(. , .)]  $ of the empirical density ${\hat p}(.) $ , of the empirical activities ${\hat a}(. , .) $ and of the empirical currents ${\hat j}(. , .) $
\begin{eqnarray}
{\cal P}^{[2.5]}_T[ {\hat p}(.); {\hat a}(. , .); {\hat j}(. , .)]  
\opsimeq_{T \to + \infty}  \delta \left(  \sum_{C} {\hat p}(C )-1\right)
\left[ \prod_{C} \delta \left( \sum_{C' \ne C} {\hat j} (C,C')  \right) \right]
  e^{ \displaystyle - T  I^{[2.5]}[ {\hat p}(.) ; {\hat a}(. , .); {\hat j}(. , .)]}
\label{proba2.5jumpPaj}
\end{eqnarray}
with the explicit rate function 
\begin{eqnarray}
&& I^{[2.5]}[ {\hat p}(.);  {\hat a}(. , .); {\hat j}(. , .)] 
 = \sum_{C}  \sum_{C' > C } 
 \bigg[ {\hat w}(C , C' ) {\hat p}(C' ) \ln \left( \frac{ {\hat w}(C, C' ) }{w(C, C' ) } \right)
- {\hat w}(C, C' ){\hat p}(C' ) + w(C , C' ){\hat p}(C' )
\nonumber \\
&& \ \ \ \ \ \ \ \ \ \ \ \ \  \ \ \ \ \ \ \ \ \ \ \ \ \  \ \ \ \ \ \ \ \ \ \ \ \ \  \ \ \ \ \ \ \ \ \ 
+   {\hat w}(C' , C ) {\hat p}(C ) \ln \left( \frac{ {\hat w}(C', C ) }{w(C' , C ) } \right)
- {\hat w}(C', C ){\hat p}(C ) + w(C' , C ){\hat p}(C )
\bigg]
\nonumber \\
&&  = \sum_{C}  \sum_{C' > C } 
 \bigg[  \frac{{\hat a} (C,C') + {\hat j} (C,C') }{ 2  } 
 \ln \left( \frac{  {\hat a} (C,C') + {\hat j} (C,C')  }{2 w(C, C' )  {\hat p}(C' )} \right) 
 +   \frac{ {\hat a} (C,C') - {\hat j} (C,C') }{ 2   }  
\ln \left( \frac{  {\hat a} (C,C') - {\hat j} (C,C')  }{2w(C' , C )  {\hat p}(C ) } \right)
\nonumber \\
&& \ \ \ \ \ \ \ \ \ \ \ \ \ 
-  {\hat a} (C,C') + w(C , C' ){\hat p}(C' ) + w(C' , C ){\hat p}(C )
\bigg]
\label{rate2.5jumpPaj}
\end{eqnarray}

For any given empirical density $ {\hat p}(.)$ and empirical activities ${\hat a}(. , .) $,
it is interesting to consider the reversal of all the empirical currents ${\hat j} (C,C') \to - {\hat j} (C,C') $ :
the divergenceless constraint is still satisfied,
while the difference between the two rate functions reduce to the following linear contribution with respect to the empirical currents ${\hat j} (C,C') $
\begin{eqnarray}
 I^{[2.5]}[ {\hat p}(.);  {\hat a}(. , .); {\hat j}(. , .)] - I^{[2.5]}[ {\hat p}(.);  {\hat a}(. , .); - {\hat j}(. , .)]
   = \sum_{C}  \sum_{C' > C }  {\hat j} (C,C') 
 \ln \left( \frac{  w(C' , C )  {\hat p}(C )  }{ w(C, C' )  {\hat p}(C' )} \right) 
\label{rate2.5jumpPajdifference}
\end{eqnarray}
This can be considered as an example of the Gallavotti-Cohen fluctuation relations
(see \cite{galla,kurchan_langevin,Leb_spo,maes1999,jepps,derrida-lecture,harris_Schu,kurchan,searles,zia,chetrite_thesis,maes2009,maes2017,chetrite_HDR} and references therein).

In the following subsections, various intermediate levels between the standard level 2.5 discussed above and the standard level 2 discussed in the previous section \ref{sec_level2} will be discussed 
with the three non-standard notations $2.25'$, $2.25$ and $2.1$ that are only meant to stress that 
the levels $2.25'$ and $2.25$ belong to two different contraction paths between 2.5 and 2, 
and that the level $2.1$ is intermediate between $2.25$ and $2$.


\subsubsection{ Explicit large deviations for the joint distribution ${\cal P}^{[2.25']}_T[ {\hat p}(.);  {\hat j}(. , .)]  $ of the empirical density and the empirical currents }

Since the activities $ {\hat a} (C,C')$ do not appear in the constitutive constraints of Eq. \ref{proba2.5jumpPaj},
one can optimize the rate function $I^{[2.5]}[ {\hat p}(.); ; {\hat a}(. , .); {\hat j}(. , .)]  $
over the activities ${\hat a}(C , C') $ 
\begin{eqnarray}
0 = \frac{ \partial I^{[2.5]}[ {\hat p}(.);  {\hat a}(. , .); {\hat j}(. , .)]}{\partial {\hat a}(C , C')} = 
 \frac{1 }{ 2  } 
 \ln \left( \frac{  {\hat a}^2 (C,C') - {\hat j}^2 (C,C')  }{4 w(C, C' )w(C' , C )  {\hat p}(C )  {\hat p}(C' )} \right) 
\label{deria}
\end{eqnarray}
to obtain their optimal values ${\hat a}^{opt}(. , .) $  as a function of the 
empirical density and empirical currents 
\begin{eqnarray}
{\hat a}^{opt}(C , C') = \sqrt{ {\hat j}^2 (C,C')  +4 w(C, C' )w(C' , C )  {\hat p}(C )  {\hat p}(C' )} 
\label{deriaopt}
\end{eqnarray}
Plugging these optimal values into Eq. \ref{rate2.5jumpPaj}
yields the explicit rate function 
\begin{small}
\begin{eqnarray}
&& I^{[2.25']}[ {\hat p}(.);  {\hat j}(. , .)]   = I^{[2.5]}[ {\hat p}(.);  {\hat a}^{opt}(. , .); {\hat j}(. , .)] 
\nonumber \\
&&  = \sum_{C}  \sum_{C' > C } 
 \bigg[  \frac{\sqrt{ {\hat j}^2 (C,C')  +4 w(C, C' )w(C' , C )  {\hat p}(C )  {\hat p}(C' )} + {\hat j} (C,C') }{ 2  } 
 \ln \left( \frac{  \sqrt{ {\hat j}^2 (C,C')  +4 w(C, C' )w(C' , C )  {\hat p}(C )  {\hat p}(C' )} + {\hat j} (C,C')  }{2 w(C, C' )  {\hat p}(C' )} \right) 
\nonumber \\
&& +   \frac{ \sqrt{ {\hat j}^2 (C,C')  +4 w(C, C' )w(C' , C )  {\hat p}(C )  {\hat p}(C' )} - {\hat j} (C,C') }{ 2   }  
\ln \left( \frac{  \sqrt{ {\hat j}^2 (C,C')  +4 w(C, C' )w(C' , C )  {\hat p}(C )  {\hat p}(C' )} - {\hat j} (C,C')  }{2w(C' , C )  {\hat p}(C ) } \right)
\nonumber \\
&& \ \ \ \ \ \ \ \ \ \ \ \ \ 
-  \sqrt{ {\hat j}^2 (C,C')  +4 w(C, C' )w(C' , C )  {\hat p}(C )  {\hat p}(C' )} + w(C , C' ){\hat p}(C' ) + w(C' , C ){\hat p}(C )
\bigg]
\label{rate2.25jumpPj}
\end{eqnarray}
\end{small}
that governs the large deviations 
of the joint distribution ${\cal P}^{[2.25']}_T[ {\hat p}(.);  {\hat j}(. , .)] $ of the empirical density ${\hat p}(.) $ 
and of the empirical currents ${\hat j}(. , .) $
\begin{eqnarray}
{\cal P}^{[2.25']}_T[ {\hat p}(.);  {\hat j}(. , .)]  
\opsimeq_{T \to + \infty}  \delta \left(  \sum_{C} {\hat p}(C )-1\right)
\left[ \prod_{C} \delta \left( \sum_{C' \ne C} {\hat j} (C,C')  \right) \right]
  e^{ \displaystyle - T  I^{[2.25']}[ {\hat p}(.) ;  {\hat j}(. , .)]}
\label{proba2.25jumpPj}
\end{eqnarray}
For an arbitrary irreversible generator, this is usually the lowest level with explicit large deviations.
In particular, the optimization of ${\cal P}^{[2.25']}_T[ {\hat p}(.);  {\hat j}(. , .)]  $ 
over the divergenceless empirical currents ${\hat j}(. , .) $ cannot be solved explicitly to obtain an explicit expression 
for the rate function 
$I^{[2]}[ {\hat p}(.) ] $ at level 2 of the empirical density ${\hat p}(.) $, in consistency with the discussions 
around Eq. \ref{legendreleftl} of the previous section.

Since in the previous section we have written the explicit form of Eq. \ref{wconditionedp}
for the generator $w^{Cond[\hat p(.)]} $ conditioned to produce the given empirical density ${\hat p}(.) $,
it is worth mentioning that one can use Eqs \ref{jempiinverse} and \ref{deriaopt}
to write the generator $w^{Cond[{\hat p}(.);  {\hat j}(. , .)]} $ conditioned to produce the given empirical density ${\hat p}(.) $ and the given empirical currents $ {\hat j}(. , .)$
\begin{eqnarray}
w^{Cond[\hat p(.);  {\hat j}(. , .)]}(C,C') 
&& = \frac{{\hat a}^{opt} (C,C') + {\hat j} (C,C') }{ 2 {\hat p}(C' ) } 
= \frac{ \sqrt{ {\hat j}^2 (C,C')  +4 w(C, C' )w(C' , C )  {\hat p}(C )  {\hat p}(C' )}  + {\hat j} (C,C') }{ 2 {\hat p}(C' ) } 
 \ \ \ {\rm for } \ \ C \ne C'
 \nonumber \\
w^{Cond[\hat p(.);  {\hat j}(. , .)]}(C',C') && = - \sum_{C \ne C'} w^{Cond[\hat p(.);  {\hat j}(. , .)]}(C,C') 
\label{wconditionedpj}
\end{eqnarray}


\subsection{ Explicit contractions from the level 2.5 towards lower levels when the generator $w(.,.)$
is reversible    }

When the generator $w$ satisfies detailed-balance, 
one can still be interested into the large deviations involving non-vanishing empirical currents ${\hat j}(. , .) \ne 0 $
as described by the explicit level 2.5 of Eq. \ref{rate2.5jumpPaj} or the explicit level 2.25 of Eq. \ref{proba2.25jumpPj},
ou into the conditioning of the reversible generator into an irreversible generator $w^{Cond[{\hat p}(.);  {\hat j}(. , .)]} $ of Eq. \ref{wconditionedpj} producing non-vanishing empirical currents ${\hat j}(. , .) \ne 0 $,
but one can also write further explicit rate functions involving vanishing empirical currents ${\hat j}(. , .) = 0 $
as described in the present subsection.


\subsubsection{ Explicit level 2.25 for joint distribution ${\cal P}^{[2.25]}_T[ {\hat p}(.);  {\hat a}(. , .)]  $ of the empirical density ${\hat p}(.)$ and the empirical activities ${\hat a}(.,.)$   }

When the generator $w$ satisfies detailed-balance, 
the Gallavotti-Cohen symmetry of Eq. \ref{rate2.5jumpPajdifference}
reduces to the symmetry upon reversal of all empirical currents
\begin{eqnarray}
 I^{[2.5]}[ {\hat p}(.);  {\hat a}(. , .); {\hat j}(. , .)] - I^{[2.5]}[ {\hat p}(.);  {\hat a}(. , .); - {\hat j}(. , .)]
   = 0
   \label{rate2.5jumpPajdifferencerev}
\end{eqnarray}
For any given empirical density ${\hat p}(.)$ and any given empirical activities ${\hat a}(.,.)$,
one then expects from a physical point of view 
that the rate function $I^{[2.5]}[ {\hat p}(.);  {\hat a}(. , .); {\hat j}(. , .)]$ is maximal for vanishing currents,
as can be checked via a short explicit calculation (see for instance the Appendix B of \cite{c_east}),
while the divergence constraint is trivially satisfied.
So one can plug these optimal vanishing values
\begin{eqnarray}
{\hat j}^{opt}(C , C') = 0
\label{joptzero}
\end{eqnarray}
into Eq. \ref{rate2.5jumpPaj}
to obtain the explicit rate function 
\begin{eqnarray}
&& I^{[2.25]}[ {\hat p}(.);  {\hat a}(. , .)]  =  I^{[2.5]}[ {\hat p}(.);  {\hat a}(. , .); {\hat j}^{opt}(. , .)=0] 
\nonumber \\
&&  = \sum_{C}  \sum_{C' > C } 
 \bigg[  \frac{{\hat a} (C,C')  }{ 2  } 
 \ln \left( \frac{  {\hat a}^2 (C,C')   }{4 w(C, C' ) w(C' , C )  {\hat p}(C ) {\hat p}(C' )} \right) 
-  {\hat a} (C,C') + w(C , C' ){\hat p}(C' ) + w(C' , C ){\hat p}(C )
\bigg]
\label{rate2.25jumpPa}
\end{eqnarray}
that governs the large deviations for the joint distribution ${\cal P}^{[2.25]}_T[ {\hat p}(.);  {\hat a}(. , .)]  $  of the empirical density ${\hat p}(.)$ and the empirical activities ${\hat a}(.,.)$
\begin{eqnarray}
{\cal P}^{[2.25]}_T[ {\hat p}(.);  {\hat a}(. , .)]  
\opsimeq_{T \to + \infty}  \delta \left(  \sum_{C} {\hat p}(C )-1\right)
  e^{ \displaystyle - T  I^{[2.25]}[ {\hat p}(.) ;  {\hat a}(. , .)]}
\label{proba2.25jumpPa}
\end{eqnarray}

The fact that all the empirical currents of Eq. \ref{jempi}
vanish
\begin{eqnarray}
0 = {\hat j} (C,C') = {\hat w}(C,C' ){\hat p}(C' ) - {\hat w}(C',C ){\hat p}(C ) 
\label{jempi0}
\end{eqnarray}
mean that the empirical transition rates ${\hat w}(. , .)$ satisfy detailed-balance with respect to the empirical density ${\hat p}(.) $, so that it is useful to introduce the empirical symmetric matrix ${\hat D}(.,.)$
to write the parametrization analog to Eq. \ref{detailedratio}
\begin{eqnarray}
{\hat w}^{DB}(C',C) = {\hat D}(C',C) \sqrt{ \frac{ {\hat p} (C')}{{\hat p} (C)} }
\label{detailedratioempi}
\end{eqnarray}
so that the empirical activities of Eq. \ref{jempi} become
\begin{eqnarray}
 {\hat a} (C,C') = {\hat w}(C,C' ){\hat p}(C' ) + {\hat w}(C',C ){\hat p}(C ) 
 = 2 {\hat D}(C',C) \sqrt{  {\hat p} (C') {\hat p} (C)} 
\label{aempi}
\end{eqnarray}

Since in the previous section we have written the explicit form of Eq. \ref{wconditionedp}
for the generator $w^{Cond[\hat p(.)]} $ conditioned to produce the given empirical density ${\hat p}(.) $,
it is worth mentioning that one can use Eqs \ref{detailedratioempi} and \ref{aempi}
to write the generator $w^{Cond[{\hat p}(.);  {\hat a}(. , .)]} $ conditioned to produce the given empirical density ${\hat p}(.) $ and the given empirical activities $ {\hat a}(. , .)$
\begin{eqnarray}
w^{Cond[\hat p(.);  {\hat a}(. , .)]}(C,C') 
&& = \frac{{\hat a} (C,C') }{ 2 {\hat p}(C' ) }  \ \ \ {\rm for } \ \ C \ne C'
 \nonumber \\
w^{Cond[\hat p(.);  {\hat a}(. , .)]}(C',C') && = - \sum_{C \ne C'} w^{Cond[\hat p(.);  {\hat a}(. , .)]}(C,C') 
= - \sum_{C \ne C'}  \frac{{\hat a} (C,C')  }{ 2 {\hat p}(C' ) }
\label{wconditionedpa}
\end{eqnarray}

One can use the parametrizations of Eqs \ref{detailedratio} and  \ref{detailedratioempi}
in order to translate Eq. \ref{proba2.25jumpPa} into the 
joint distribution ${\cal P}^{[2.25]}_T[ {\hat p}(.); {\hat D}(. , .)] $ 
of the empirical density ${\hat p}(.)$ and of the empirical symmetric matrix ${\hat D}(.,.)$
\begin{eqnarray}
{\cal P}^{[2.25]}_T[ {\hat p}(.);  {\hat D}(. , .)]  
\opsimeq_{T \to + \infty}  \delta \left(  \sum_{C} {\hat p}(C )-1\right)
  e^{ \displaystyle - T  I^{[2.25]}[ {\hat p}(.) ;  {\hat D}(. , .)]}
\label{proba2.25jumpPD}
\end{eqnarray}
with the explicit rate function
\begin{eqnarray}
 I^{[2.25]}[ {\hat p}(.);  {\hat D}(. , .)] 
&&  = \sum_{C}  \sum_{C' > C } 
 \bigg[ 2 {\hat D}(C',C) \sqrt{  {\hat p} (C') {\hat p} (C)}   
 \ln \left( \frac{ {\hat D}(C',C)     }{ D(C' , C )  } \right) 
 - 2 {\hat D}(C',C) \sqrt{  {\hat p} (C') {\hat p} (C)}
\nonumber \\
&& \ \ \ \ \ \ \ \ \ \ \ \ \ 
  + D(C',C) \sqrt{ \frac{ p_* (C)}{p_* (C')} }{\hat p}(C' ) 
+ D(C',C) \sqrt{ \frac{ p_* (C')}{p_* (C)} }{\hat p}(C )
\bigg]
\nonumber \\
&& =\sum_{C}  \sum_{C' > C }  \sqrt{  {\hat p} (C') {\hat p} (C)} 
\bigg[ 2 {\hat D}(C',C) 
 \ln \left( \frac{ {\hat D}(C',C)  }{ D(C',C)  } \right)
- 2 {\hat D}(C',C)  
\nonumber \\ && \ \ \ \ \ \ \ \ \ 
 +  D(C',C) \sqrt{ \frac{ p_* (C)}{p_* (C')} }\sqrt{ \frac{ {\hat p} (C')}{{\hat p} (C)} }
+  D(C',C) \sqrt{ \frac{ p_* (C')}{p_* (C)} }\sqrt{ \frac{ {\hat p} (C)}{{\hat p} (C')} }
\bigg]
\label{rate2.25jumpPD}
\end{eqnarray}


\subsubsection{ Recovering the explicit level 2 for the empirical density ${\hat p}(.)$  via contractions of higher levels }

Within the present perspective, the large deviations at level 2 for the empirical density ${\hat p}(.)$ discussed in 
the previous section \ref{sec_level2}
can be recovered 
via two possible contractions :

(i) The optimization of ${\cal P}^{[2.25]}_T[ {\hat p}(.);  {\hat j}(. , .)]  $ of Eq. \ref{proba2.25jumpPa}
over the empirical currents ${\hat j}(. , .) $ corresponds to vanishing currents ${\hat j}^{opt}(. , .)=0 $ as in Eq. \ref{joptzero} and leads to the rate function at level 2 
\begin{eqnarray}
 I^{[2]}[ {\hat p}(.)]&& = I^{[2.25]}[ {\hat p}(.);  {\hat j}^{opt}(. , .)=0]  
\nonumber \\ &&  = \sum_{C}  \sum_{C' > C } 
 \bigg[  
-  \sqrt{ 4 w(C, C' )w(C' , C )  {\hat p}(C )  {\hat p}(C' )} + w(C , C' ){\hat p}(C' ) + w(C' , C ){\hat p}(C )
\bigg]
\nonumber \\ &&  = \sum_{C}  \sum_{C' > C } 
 \bigg[  
  \sqrt{ w(C' , C ){\hat p}(C )} - \sqrt{ w(C , C' ){\hat p}(C' ) } 
\bigg]^2
\label{rate2from2.25jumpPa}
\end{eqnarray}
where one can plug the detailed-balance parametrization of Eq. \ref{detailedratio}
for the reversible generator $w$ to recover Eq. \ref{I2assumofsquares}
\begin{eqnarray}
 I^{[2]}[ {\hat p}(.)]&&   = \sum_{C}  \sum_{C' > C } D(C',C)  \sqrt{p_* (C') p_* (C) }
 \bigg[  
  \sqrt{   \frac{ {\hat p}(C )}{ p_* (C)} } -  \sqrt{ \frac{ {\hat p}(C' )}{p_* (C')} } 
\bigg]^2
\label{rate2from2.25jumpPaD}
\end{eqnarray}

(ii) The optimization of the rate function $I^{[2.25]}[ {\hat p}(.); {\hat D}(. , .)] $ of Eq. \ref{rate2.25jumpPD}
over the empirical matrix elements ${\hat D}(.)$
\begin{eqnarray}
0 = \frac{ \partial I^{[2.25]}[ {\hat p}(.); {\hat D}(. , .)] }{ \partial {\hat D}(C',C)}
 = 2 \sqrt{  {\hat p} (C') {\hat p} (C)} 
 \ln \left( \frac{ {\hat D}(C',C)  }{ D(C',C)  } \right)
\label{rate2.25jumpPwderi}
\end{eqnarray}
yields that the optimal values ${\hat D}^{opt}(C',C) $ coincide with the true matrix elements $D(C',C)$
\begin{eqnarray}
{\hat D}^{opt}(C',C) = D(C',C)
\label{empiDoptimal}
\end{eqnarray}
So the rate function at level 2 for the empirical density ${\hat p}(.)$ 
can be also found by plugging this optimal value into the rate function of Eq. \ref{rate2.25jumpPD}
\begin{eqnarray}
 I^{[2]}[{\hat p}(.)   ] && =I^{[2.25]}[ {\hat p}(.); {\hat D}^{opt}(. , .)=D(.,.)] 
\nonumber \\
&&  = \sum_{C}  \sum_{C' > C }  \sqrt{  {\hat p} (C') {\hat p} (C)} 
\bigg[ 
- 2 D (C',C)  +  D(C',C) \sqrt{ \frac{ p_* (C')}{p_* (C)} }\sqrt{ \frac{ {\hat p} (C)}{{\hat p} (C')} }
 +  D(C',C) \sqrt{ \frac{ p_* (C)}{p_* (C')} }\sqrt{ \frac{ {\hat p} (C')}{{\hat p} (C)} }
\bigg]
\nonumber \\
&&  = \sum_{C}  \sum_{C' > C }  D (C',C)  \sqrt{p_* (C') p_* (C) } 
\bigg[ 
- 2  \frac{ \sqrt{  {\hat p} (C') {\hat p} (C)} }{ \sqrt{p_* (C') p_* (C) } } 
 +    \frac{ {\hat p} (C)}{p_* (C) }
 +   \frac{ {\hat p} (C')}{p_* (C') }
\bigg]
\nonumber \\
&&  = \sum_{C}  \sum_{C' > C }  D (C',C)  \sqrt{p_* (C') p_* (C) } 
\bigg[ 
 \sqrt{    \frac{ {\hat p} (C)}{p_* (C) } }
 -   \sqrt{    \frac{ {\hat p} (C')}{p_* (C') } }
\bigg]^2
 \label{rate2.jumpPw}
\end{eqnarray}
to recover Eq. \ref{I2assumofsquares} and \ref{rate2from2.25jumpPaD}.


\subsubsection{ Explicit level 2.1 for the joint distribution ${\cal P}^{[2.1]}_T[ {\hat p}(.);  {\hat A}]  $ of the empirical density ${\hat p}(.)$ and the total empirical activity ${\hat A}$   }

The total empirical activity ${\hat A}(.,.)$ is defined as the sum of the local activities ${\hat a} (C,C') $ of Eq. \ref{aempi}
over all the links
\begin{eqnarray}
{\hat A} && \equiv \sum_{C } \sum_{C' > C} {\hat a} (C,C') 
=\sum_{C } \sum_{C' > C} \left[  {\hat w}(C,C' ){\hat p}(C' ) + {\hat w}(C',C ){\hat p}(C )   \right]
\nonumber \\
&& = \sum_{C } \sum_{C' > C} 2 {\hat D}(C',C) \sqrt{  {\hat p} (C') {\hat p} (C)}
\label{Atot}
\end{eqnarray}
In terms of the trajectory $C(0 \leq t \leq T)$ over the large-time window $t \in [0,T]$, 
Eqs \ref{Qempi} and \ref{jempi} yield that the total empirical activity ${\hat A} $ multiplied by $T$
\begin{eqnarray}
T {\hat A} && 
=\sum_{C } \sum_{C' > C} \left[ 
\sum_{t \in [0,T] :C(t^+) \ne C(t) }
\left(\delta_{C(t^+) ,C'} \delta_{C(t) ,C}  +\delta_{C(t^+) ,C'} \delta_{C(t) ,C} \right) \right]
\nonumber \\
&& =  \sum_{t \in [0,T] :C(t^+) \ne C(t) } 1 \equiv  {\hat N}_T
\label{Atottraj}
\end{eqnarray}
simply represents the total number $N_T$ of jumps 
observed along the trajectory $C(0 \leq t \leq T)$ during the time-window $[0,T]$.
This observable has attracted a lot of interest in the field of reversible glassy dynamics,
in particular for Kinetically-Constrained-Models
(see the reviews \cite{ritort2003,lecture,chapter,math2013,garrahan_lecture} and references therein),
and its large deviations properties have been much studied in various models \cite{garrahan_lecture,lecomte_glass,kristina1,kristina2,bodineau,LeeYang,jack_soft,jack_sol,c_east,banuls,noninter,XOR,MPS,garrahan2021,sollich_MFtrap}.

The joint distribution ${\cal P}^{[2.1]}_T[ {\hat p}(.);  {\hat A}]  $ of the empirical density ${\hat p}(.)$ and the total empirical activity ${\hat A}(.,.)$ can be obtained from the integration of Eq. \ref{proba2.25jumpPD}
over ${\hat D}(.,.) $ in the presence of the constraint that imposes the correct value for the total activity of Eq. \ref{Atot}
\begin{eqnarray}
{\cal P}^{[2.1]}_T[ {\hat p}(.);  {\hat A}] && = \int d {\hat D}(. , .){\cal P}^{[2.25]}_T[ {\hat p}(.);  {\hat D}(. , .)]  
\delta \left(\sum_{C } \sum_{C' > C} 2 {\hat D}(C',C) \sqrt{  {\hat p} (C') {\hat p} (C)} - {\hat A} \right)
\nonumber \\
&&\opsimeq_{T \to + \infty}  \delta \left(  \sum_{C} {\hat p}(C )-1\right)
\int d {\hat D}(. , .)  e^{ \displaystyle - T  I^{[2.25]}[ {\hat p}(.) ;  {\hat D}(. , .)]}
\delta \left(\sum_{C } \sum_{C' > C} 2 {\hat D}(C',C) \sqrt{  {\hat p} (C') {\hat p} (C)} - {\hat A} \right)
\nonumber \\
&&\opsimeq_{T \to + \infty}\delta \left(  \sum_{C} {\hat p}(C )-1\right)
  e^{ \displaystyle - T  I^{[2.1]}[ {\hat p}(.) ;  {\hat A}]}
\label{proba2.1jumpPAtot}
\end{eqnarray}
So the rate function $I^{[2.1]}[ {\hat p}(.) ;  {\hat A}] $ corresponds to the optimization of the rate function $I^{[2.25]}[ {\hat p}(.) ;  {\hat D}(. , .)] $ over ${\hat D}(. , .) $ in the presence of the constraint that can be taken into account via
the Lagrange multiplier $\lambda$ in the Lagrangian
\begin{eqnarray}
&& {\cal L}[ {\hat D}(. , .)] =  I^{[2.25]}[ {\hat p}(.);  {\hat D}(. , .)] 
+ \lambda \left(\sum_{C } \sum_{C' > C} 2 {\hat D}(C',C) \sqrt{  {\hat p} (C') {\hat p} (C)} - {\hat A} \right)
 \nonumber \\
 &&   =\sum_{C}  \sum_{C' > C }  \sqrt{  {\hat p} (C') {\hat p} (C)} 
\bigg[ 2 {\hat D}(C',C) 
 \ln \left( \frac{ {\hat D}(C',C)  }{ D(C',C)  } \right)
- 2 {\hat D}(C',C)  
 +  D(C',C) \sqrt{ \frac{ p_* (C)}{p_* (C')} }\sqrt{ \frac{ {\hat p} (C')}{{\hat p} (C)} }
+  D(C',C) \sqrt{ \frac{ p_* (C')}{p_* (C)} }\sqrt{ \frac{ {\hat p} (C)}{{\hat p} (C')} }
\bigg]
 \nonumber \\ && 
 + \lambda \left(\sum_{C } \sum_{C' > C} 2 {\hat D}(C',C) \sqrt{  {\hat p} (C') {\hat p} (C)} - {\hat A} \right)
\label{rate2.25jumpPDlagrangian}
\end{eqnarray}

The optimization of this Lagrangian
over the empirical matrix elements ${\hat D}(.)$
\begin{eqnarray}
0 = \frac{ \partial  {\cal L}[ {\hat D}(. , .)]  }{ \partial {\hat D}(C',C)}
 = 2 \sqrt{  {\hat p} (C') {\hat p} (C)} \left[  \ln \left( \frac{ {\hat D}(C',C)  }{ D(C',C)  } \right) + \lambda \right]
\label{rate2.25jumpPwderiD}
\end{eqnarray}
yields that the optimal values ${\hat D}^{opt}(C',C) $ are simply proportional to the true matrix elements $D(C',C)$
\begin{eqnarray}
{\hat D}^{opt}(C',C) = D(C',C) e^{- \lambda}
\label{empiDoptimalD}
\end{eqnarray}
where the common factor $ e^{- \lambda}$ is determined by the satisfaction of the constraint
\begin{eqnarray}
{\hat A}  = \sum_{C } \sum_{C' > C} 2 {\hat D}^{opt}(C',C) \sqrt{  {\hat p} (C') {\hat p} (C)}
=e^{- \lambda} \sum_{C } \sum_{C' > C} 2 D^{opt}(C',C) \sqrt{  {\hat p} (C') {\hat p} (C)}
\equiv e^{- \lambda} {\cal A} [ {\hat p}(.)] 
\label{Atotlambda}
\end{eqnarray}
where we have introduced the notation 
\begin{eqnarray}
{\cal A} [ {\hat p}(.)]  \equiv \sum_{C } \sum_{C' > C} 2 D(C',C) \sqrt{  {\hat p} (C') {\hat p} (C)}
\label{calAp}
\end{eqnarray}
for the total activity that would be produced by the empirical density ${\hat p}(.) $ in the presence of the true matrix elements $D(C',C)$.
Plugging the optimal values 
\begin{eqnarray}
{\hat D}^{opt}(C',C) = D(C',C) e^{- \lambda} = D(C',C) \frac{{\hat A} }{{\cal A} [ {\hat p}(.)] } 
\label{empiDoptimalDrep}
\end{eqnarray}
into the rate function $I^{[2.25]}[ {\hat p}(.) ;  {\hat D}(. , .)] $ of Eq. \ref{rate2.25jumpPD}
yields the explicit rate function 
\begin{eqnarray}
 I^{[2.1]}[ {\hat p}(.) ;  {\hat A}]
 && =  I^{[2.25]}[ {\hat p}(.);  {\hat D}^{opt}(. , .)] 
 \nonumber \\
 &&   =\sum_{C}  \sum_{C' > C }  \sqrt{  {\hat p} (C') {\hat p} (C)} 
\bigg[ 2 D(C',C) \frac{{\hat A} }{{\cal A} [ {\hat p}(.)] } 
 \ln \left(  \frac{{\hat A} }{{\cal A} [ {\hat p}(.)] }    \right)
- 2 D(C',C) \frac{{\hat A} }{{\cal A} [ {\hat p}(.)] }  
 \nonumber \\ && \ \ \ \ \ \ \ \ \ 
 +  D(C',C) \sqrt{ \frac{ p_* (C)}{p_* (C')} }\sqrt{ \frac{ {\hat p} (C')}{{\hat p} (C)} }
+  D(C',C) \sqrt{ \frac{ p_* (C')}{p_* (C)} }\sqrt{ \frac{ {\hat p} (C)}{{\hat p} (C')} }
\bigg]
 \nonumber \\
 &&   = {\hat A} \ln \left(  \frac{{\hat A} }{{\cal A} [ {\hat p}(.)] }    \right) - {\hat A}
+ \sum_{C}  \sum_{C' > C }  D(C',C) \sqrt{  {\hat p} (C') {\hat p} (C)} 
\bigg[ 
   \sqrt{ \frac{ p_* (C)}{p_* (C')} }\sqrt{ \frac{ {\hat p} (C')}{{\hat p} (C)} }
+   \sqrt{ \frac{ p_* (C')}{p_* (C)} }\sqrt{ \frac{ {\hat p} (C)}{{\hat p} (C')} }
\bigg]
\label{rate2.1jumpPA}
\end{eqnarray}

In particular, the difference with the rate function $ I^{[2]}[ {\hat p}(.) ] $ of Eq. \ref{rate2.jumpPw}
concerning the empirical density ${\hat p} (.)$ alone reduces to the simple expression
\begin{eqnarray}
 I^{[2.1]}[ {\hat p}(.) ;  {\hat A}] - I^{[2]}[ {\hat p}(.) ]
   = {\hat A} \ln \left(  \frac{{\hat A} }{{\cal A} [ {\hat p}(.)] }    \right) - {\hat A} + {\cal A} [ {\hat p}(.)]
\label{rate2.1jumpPAdifference}
\end{eqnarray}
that represents the additional cost for having an empirical activity ${\hat A} $
different from the activity ${\cal A} [ {\hat p}(.)] $ of Eq. \ref{calAp}.
Since one recognizes the rate function associated to Poisson processes,
the physical meaning is that once the empirical density ${\hat p}(.) $ is given,
the conditional probability ${\cal P}( {\hat N}_T=T {\hat A} \vert {\hat p}(.)) $
 to see the total number of jumps ${\hat N}_T = T {\hat A}$ of 
Eq. \ref{Atottraj} can be considered as the following Poisson distribution of average $N_T^{av} = T {\cal A} [ {\hat p}(.)] $
\begin{eqnarray}
{\cal P}( {\hat N}_T=T {\hat A} \vert {\hat p}(.))  = \frac{ (N_T^{av} )^{{\hat N}_T}  e^{- N_T^{av}} }{{\hat N}_T !}
&& = \frac{ (T {\cal A} [ {\hat p}(.)] )^{T {\hat A}}  e^{- T {\cal A} [ {\hat p}(.)] }}{ (T {\hat A}) !}
\nonumber \\
&& \opsimeq_{T \to + \infty} e^{ - T \left[ {\hat A} \ln \left(  \frac{{\hat A} }{{\cal A} [ {\hat p}(.)] }    \right) - {\hat A} + {\cal A} [ {\hat p}(.)]\right] }
\label{Poisson}
\end{eqnarray}
where the Stirling approximation was used for the factorial to obtain that the rate function for large $T$
coincides with Eq. \ref{rate2.1jumpPAdifference}.
Another way to characterize this Poisson statistics of the total empirical activity ${\cal A} $ once 
the empirical density ${\hat p}(.) $ is given, is by considering
 the generating function with respect to ${\cal A} $ of the joint distribution
${\cal P}^{[2.1]}_T[ {\hat p}(.);  {\hat A}] $ of Eq. \ref{proba2.1jumpPAtot}
using Eq. \ref{rate2.1jumpPAdifference}
\begin{eqnarray}
{\cal Z}^{[2.1]}_T[ {\hat p}(.); \lambda ] && \equiv \int_0^{+\infty} d {\hat A} e^{- T \lambda {\hat A}} {\cal P}^{[2.1]}_T[ {\hat p}(.);  {\hat A}]  
\opsimeq_{T \to + \infty}\delta \left(  \sum_{C} {\hat p}(C )-1\right)
\int_0^{+\infty} d {\hat A}   e^{ \displaystyle - T \lambda {\hat A} - T  I^{[2.1]}[ {\hat p}(.) ;  {\hat A}]}
\nonumber \\
&&\opsimeq_{T \to + \infty}
\delta \left(  \sum_{C} {\hat p}(C )-1\right)e^{ \displaystyle - T  I^{[2]}[ {\hat p}(.) ]}
\int_0^{+\infty} d {\hat A} e^{- T \left[ \lambda {\hat A}
+  {\hat A} \ln \left(  \frac{{\hat A} }{{\cal A} [ {\hat p}(.)] }    \right) - {\hat A} + {\cal A} [ {\hat p}(.)] \right] }
\label{geneAproba2.1calcul}
\end{eqnarray}
The saddle-point evaluation for large $T$ leads to the optimal value ${\hat A}^{opt} = e^{-\lambda }  {\cal A} [ {\hat p}(.)] $
and to the final result
\begin{eqnarray}
{\cal Z}^{[2.1]}_T[ {\hat p}(.); \lambda ] &&\opsimeq_{T \to + \infty}
\delta \left(  \sum_{C} {\hat p}(C )-1\right)e^{ \displaystyle - T \left[  I^{[2]}[ {\hat p}(.) ] + (1-e^{-\lambda}){\cal A} [ {\hat p}(.)]  \right] }
\label{geneAproba2.1}
\end{eqnarray}

It is interesting to rewrite Eq. \ref{calAp} in terms of the off-diagonal matrix elements $H(C',C)=-D(C,C') $ of Eq. \ref{offdiag}
of the supersymmetric Hamiltonian $H$
\begin{eqnarray}
{\cal A} [ {\hat p}(.)]  = \sum_{C } \sum_{C' \ne C} D (C',C)  \sqrt{  {\hat p} (C') {\hat p} (C)}
= - \sum_{C } \sum_{C' \ne C} H (C',C)   \sqrt{  {\hat p} (C') {\hat p} (C)} =  \langle  \sqrt{  {\hat p} } \vert (-H^{off} ) \vert  \sqrt{  {\hat p} } \rangle
\label{calApHoff}
\end{eqnarray}
while the rate function at level 2 of Eq. \ref{legendreleftrightexplicit} 
reads in terms of the matrix elements of the diagonal part $H^{diag} $ and of the off-diagonal part $H^{off} $ of the Hamiltonian $H$
\begin{eqnarray}
I^{[2]}  [ {\hat p}(.)]
 = \langle \sqrt{ {\hat p}} \vert H \vert \sqrt{ {\hat p}} \rangle
 = \langle \sqrt{ {\hat p}} \vert H^{diag} \vert \sqrt{ {\hat p}} \rangle + \langle \sqrt{ {\hat p}} \vert H^{off}  \vert \sqrt{ {\hat p}} \rangle
\label{legendreleftrightexplicitdiagetoff}
\end{eqnarray}

In conclusion, the rate function $  I^{[2.1]}[ {\hat p}(.) ;  {\hat A}] $ of Eq. \ref{rate2.1jumpPAdifference} can be rewritten 
in terms of the two matrix elements $\langle \sqrt{ {\hat p}} \vert H^{diag} \vert \sqrt{ {\hat p}} \rangle $ and $\langle \sqrt{ {\hat p}} \vert H^{off} \vert \sqrt{ {\hat p}} \rangle $
\begin{eqnarray}
 I^{[2.1]}[ {\hat p}(.) ;  {\hat A}] && = I^{[2]}[ {\hat p}(.) ]
   + {\hat A} \ln \left(  \frac{{\hat A} }{{\cal A} [ {\hat p}(.)] }    \right) - {\hat A} + {\cal A} [ {\hat p}(.)]
   \nonumber \\
   && = \langle \sqrt{ {\hat p}} \vert H^{diag} \vert \sqrt{ {\hat p}} \rangle 
   + {\hat A} \ln \left(  \frac{{\hat A} }{\langle  \sqrt{  {\hat p} } \vert (-H^{off} ) \vert  \sqrt{  {\hat p} } \rangle }    \right) - {\hat A} 
   \label{rate2.1jumpPAH}
\end{eqnarray}
while the generating function of Eq. \ref{geneAproba2.1} becomes
\begin{eqnarray}
{\cal Z}^{[2.1]}_T[ {\hat p}(.); \lambda ] &&\opsimeq_{T \to + \infty}
\delta \left(  \sum_{C} {\hat p}(C )-1\right)e^{ \displaystyle - T \left[   \langle \sqrt{ {\hat p}} \vert H^{diag} \vert \sqrt{ {\hat p}} \rangle 
 + e^{-\lambda}\langle  \sqrt{  {\hat p} } \vert H^{off}  \vert  \sqrt{  {\hat p} } \rangle  \right] }
 \nonumber \\
 && \opsimeq_{T \to + \infty}
\delta \left(  \sum_{C} {\hat p}(C )-1\right)e^{ \displaystyle - T \left[   \langle \sqrt{ {\hat p}} \vert \bigg( H^{diag} 
+ e^{-\lambda}H^{off} \bigg) \vert \sqrt{ {\hat p}} \rangle    \right] }
\label{geneAproba2.1H}
\end{eqnarray}

Since we have written the explicit form of Eq. \ref{wconditionedp}
for the generator $w^{Cond[\hat p(.)]} $ conditioned to produce the given empirical density ${\hat p}(.) $,
as well as the explicit form of Eq. \ref{wconditionedpa}
the generator $w^{Cond[{\hat p}(.);  {\hat a}(. , .)]} $ conditioned to produce the given empirical density ${\hat p}(.) $ and the given empirical activities $ {\hat a}(. , .)$,
it is worth mentioning that one can use Eq \ref{empiDoptimalDrep} 
to write the generator $w^{Cond[{\hat p}(.);  {\hat A}]} $ conditioned to produce the given empirical density ${\hat p}(.) $ and the given total empirical activity $ {\hat A}$ 
in terms of the function ${\cal A} [ {\hat p}(.)] $ of Eq. \ref{calApHoff}
\begin{eqnarray}
w^{Cond[\hat p(.);  {\hat A}]}(C,C') 
&& =   D(C , C') \frac{{\hat A} }{{\cal A} [ {\hat p}(.)] }  \sqrt{\frac{ {\hat p}(C)}{ {\hat p} (C')} }  
  \ \ \ {\rm for } \ \ C \ne C'
 \nonumber \\
w^{Cond[\hat p(.);  {\hat A}]}(C',C') && = - \sum_{C \ne C'} w^{Cond[\hat p(.);  {\hat A}]}(C,C') 
\label{wconditionedpatot}
\end{eqnarray}


\subsubsection{ Link with the large deviations of the total empirical activity ${\hat A}$ alone  }

If one is interested into the distribution ${\cal P}_T[ {\hat A}] $ of the empirical activity ${\hat A} $ alone,
one can integrate the joint distribution ${\cal P}^{[2.1]}_T[ {\hat p}(.);  {\hat A}]  $ of Eq. \ref{proba2.1jumpPAtot}
over the empirical density ${\hat p}(.)$ 
\begin{eqnarray}
{\cal P}_T[ {\hat A}] && = \int d {\hat p}(.) {\cal P}^{[2.1]}_T[ {\hat p}(.);  {\hat A}] 
\nonumber \\
&&\opsimeq_{T \to + \infty} \int d {\hat p}(.) \delta \left(  \sum_{C} {\hat p}(C )-1\right)
  e^{ \displaystyle - T  I^{[2.1]}[ {\hat p}(.) ;  {\hat A}]} \opsimeq_{T \to + \infty} e^{ \displaystyle - T  I[ {\hat A}]}
\label{probaAtot}
\end{eqnarray}
while its generating function ${\cal Z}_T[  \lambda ] $ corresponds to the integration of ${\cal Z}^{[2.1]}_T[ {\hat p}(.); \lambda ]  $
of Eqs \ref{geneAproba2.1calcul} \ref{geneAproba2.1H} over the empirical density ${\hat p}(.)$ 
\begin{eqnarray}
{\cal Z}_T[  \lambda ] 
&& \equiv \int_0^{+\infty} d {\hat A} e^{- T \lambda {\hat A}} {\cal P}_T[ {\hat A}]
= \int d {\hat p}(.){\cal Z}^{[2.1]}_T[ {\hat p}(.); \lambda ] 
 \nonumber \\
 && \opsimeq_{T \to + \infty} \int d {\hat p}(.)
\delta \left(  \sum_{C} \left[ \sqrt{ {\hat p}} (C ) \right]^2-1\right)e^{ \displaystyle - T \left[   \langle \sqrt{ {\hat p}} \vert \bigg( H^{diag} 
+ e^{-\lambda}H^{off} \bigg) \vert \sqrt{ {\hat p}} \rangle    \right] } \opsimeq_{T \to + \infty} e^{ \displaystyle - T  E( \lambda) }
\label{geneA}
\end{eqnarray}
For large $T$, one needs to minimize the energy $\langle \sqrt{ {\hat p}} \vert \bigg( H^{diag} 
+ e^{-\lambda}H^{off} \bigg) \vert \sqrt{ {\hat p}} \rangle $ of the deformed quantum Hamiltonian 
\begin{eqnarray}
H^{[\lambda]} \equiv H^{diag} + e^{-\lambda}H^{off}
\label{Hdeformedlambda}
\end{eqnarray}
for the normalized ket $\vert \sqrt{ {\hat p}} \rangle $ : one recognizes
 the well-known variational definition of the ground state energy $E_0(\lambda)$ associated to the ground state $\psi_0^{[\lambda]}(C)$
for the deformed quantum Hamiltonian $H^{[\lambda]}$
\begin{eqnarray}
E(\lambda) \vert \psi_0^{[\lambda]} \rangle = H^{[\lambda]} \vert \psi_0^{[\lambda]} \rangle 
= \bigg( H^{diag} + e^{-\lambda}H^{off} \bigg) \vert \psi_0^{[\lambda]} \rangle
\label{Hdeformeigen}
\end{eqnarray}
As it should, one thus recovers the standard method to study the large deviations of the empirical activity ${\hat A} $ alone,
based on the eigenvalue problem of Eq. \ref{Hdeformeigen}
for the Hamiltonian of Eq. \ref{Hdeformedlambda} where the off-diagonal part $H^{off} $ is deformed by the coefficient $e^{-\lambda}$, but this eigenvalue problem cannot be solved explicitly in general.


\section{ Application to pure or random spin chains with local transition rates   }

\label{sec_spins}

The goal of this section is to show that for many-body models with local transition rates,
the quantum density matrix ${\hat \rho} = \vert \sqrt{{\hat p}} \rangle \langle \sqrt{ {\hat p} } \vert $
introduced in Eq. \ref{rhotot} is useful in order to rewrite the various rate functions in terms of reduced density matrices involving only a few neighboring degrees of freedom.
For concreteness, we will focus on pure or random spin chains 
spin chains with single-spin-flip or two-spin-flip transition rates.

\subsection{ Two-spin-flip dynamics of spin chains with factorized steady states  }

\label{sec_2spinflip}

In this subsection, we consider the simple example of a spin chain of $N$ spins with periodic boundary conditions
where the steady state is factorized over the sites
\begin{eqnarray}
P_*(C=\{S_1,S_2,..,S_N\}) && = \prod_{i=1}^N p^{[i]}_*(S_i)
\nonumber \\
p^{[i]}_*(S_i=\pm 1) && =\frac{ e^{ \beta h_i S_i} }{e^{ \beta h_i } +e^{ - \beta h_i } }
\label{steadyfactorized}
\end{eqnarray}
while the steady state $p^{[i]}_*(S_i=\pm 1)$ for the spin $S_i=\pm 1$ is  parametrized by field $h_i$ that may depend on $i$.
We focus on the local two-spin-flip dynamics 
that can be interpreted as reversible exclusion processes with dimer evaporation and dimer deposition  \cite{Harris_twospinflip}, 
and also appear in the Domain-Wall formulation of the Glauber single-spin-flip dynamics of the Ising chain \cite{JackSollich2010,GlauberIsing}.


\subsubsection{  Supersymmetric quantum Hamiltonian $H$ associated to the two-spin-flip dynamics  }

For a given spin $S_i=\pm$, the local Hilbert space $\vert S_i =\pm  \rangle$ is of dimension two, 
so the operators acting on this Hilbert space correspond to $2 \times 2$ matrices.
The standard basis used in quantum mechanics is given by the four Pauli matrices $\sigma_i^{a=0,x,y,z}$, 
but for our present purposes it will be more convenient to work directly with the four matrix elements of  
$2 \times 2$ matrices, i.e. with the two projectors for the diagonal part
\begin{eqnarray}
\pi_i^+ && \equiv \vert (+)_i  \rangle \langle (+)_i  \vert = \frac{1+\sigma^z_i}{2}
\nonumber \\
\pi_i^- && \equiv \vert (-)_i  \rangle \langle (-)_i  \vert = \frac{1-\sigma^z_i}{2}
\label{pauliprojectors}
\end{eqnarray}
and with the two ladder operators for the off-diagonal part
\begin{eqnarray}
\sigma_i^+ && \equiv \vert (+)_i  \rangle \langle (-)_i  \vert = \frac{\sigma^x_i+i\sigma^y_i}{2}
\nonumber \\
\sigma_i^- && \equiv \vert (-)_i  \rangle \langle (+)_i  \vert = \frac{\sigma^x_i-i\sigma^y_i}{2}
\label{pauliladder}
\end{eqnarray}

We focus on the two-spin-flip dynamics of two neighboring spins $(S_i,S_{i+1}) \rightarrow (-S_i,-S_{i+1})$
while the other $(N-2)$ spins $S_{j \ne (i,i+1)}$ remain unchanged,
so the off-diagonal part of the quantum Hamiltonian of Eq. \ref{offdiag}
can be written as
\begin{eqnarray}
H^{off} = - \sum_{i=1}^N \left[ D_{i+\frac{1}{2}}^{F} \left(\sigma_i^+ \sigma_{i+1}^+ + \sigma_i^- \sigma_{i+1}^- \right) 
+ D_{i+\frac{1}{2}}^{AF} \left(\sigma_i^+ \sigma_{i+1}^- + \sigma_i^- \sigma_{i+1}^+ \right)\right]
\label{Hoff2spinflip}
\end{eqnarray}
where the coefficients $D_{i+\frac{1}{2}}^{F} \geq 0$ parametrize the two-spin-flip rates  
of $(S_i,S_{i+1})$ between the two ferromagnetic states $(++ \leftrightarrow --)$ via Eq. \ref{detailedratio}
\begin{eqnarray}
w[(+)_i (+)_{i+1} \leftarrow (-)_i (-)_{i+1}] && 
= D_{i+\frac{1}{2}}^{F}  \sqrt{ \frac{ p^{[i]}_*(+) p^{[i+1]}_*(+)}{ p^{[i]}_*(-) p^{[i+1]}_*(-)} }
= D_{i+\frac{1}{2}}^{F} e^{ \beta ( h_i+h_{i+1} )}
\nonumber \\
w[(-)_i (-)_{i+1} \leftarrow (+)_i (+)_{i+1}]  && 
= D_{i+\frac{1}{2}}^{F}  \sqrt{ \frac{ p^{[i]}_*(-) p^{[i+1]}_*(-)}{ p^{[i]}_*(+) p^{[i+1]}_*(+)} }
= D_{i+\frac{1}{2}}^{F} e^{- \beta ( h_i+h_{i+1} )}
\label{detailedratioferro}
\end{eqnarray}
while
the coefficients $D_{i+\frac{1}{2}}^{AF} \geq 0$ parametrize the two-spin-flip rates of $(S_i,S_{i+1})$
 between the two antiferromagnetic states $(+- \leftrightarrow -+)$
 \begin{eqnarray}
w[(+)_i (-)_{i+1} \leftarrow (-)_i (+)_{i+1}]  && 
= D_{i+\frac{1}{2}}^{AF}  \sqrt{ \frac{ p^{[i]}_*(+) p^{[i+1]}_*(-)}{ p^{[i]}_*(-) p^{[i+1]}_*(+)} }
= D_{i+\frac{1}{2}}^{AF} e^{ \beta ( h_i-h_{i+1} )}
\nonumber \\
w[(-)_i (+)_{i+1} \leftarrow (+)_i (-)_{i+1}]  && 
= D_{i+\frac{1}{2}}^{AF}  \sqrt{ \frac{ p^{[i]}_*(-) p^{[i+1]}_*(+)}{ p^{[i]}_*(+) p^{[i+1]}_*(-)} }
= D_{i+\frac{1}{2}}^{AF} e^{ \beta ( - h_i+h_{i+1} )}
\label{detailedratioantiferro}
\end{eqnarray}
 
 The diagonal part of Eq. \ref{diag}
 then reads 
\begin{eqnarray}
H^{diag} && = \sum_{i=1}^N \sum_{S_i=\pm 1} \sum_{S_{i+1}=\pm 1} 
w[ -S_i,-S_{i+1}  \leftarrow S_i,S_{i+1}] \pi_i^{S_i}  \pi_{i+1}^{S_{i+1}}
\nonumber \\
&& =  \sum_{i=1}^N \left[ D_{i+\frac{1}{2}}^{F} 
\left( e^{ \beta (h_i +h_{i+1})}\pi_i^-  \pi_{i+1}^- +e^{-  \beta (h_i +h_{i+1})} \pi_i^+  \pi_{i+1}^+\right) 
+ D_{i+\frac{1}{2}}^{AF} \left(e^{ \beta (h_i -h_{i+1})} \pi_i^-  \pi_{i+1}^+ + e^{ \beta (-h_i +h_{i+1})} \pi_i^+  \pi_{i+1}^- \right)\right] \nonumber \\
\label{Hdiag2spinflip}
\end{eqnarray}
Finally, the full Hamiltonian $H$ can be written in the supersymmetric form of Eqs \ref{Hsum} \ref{Hlinkproj}
\begin{eqnarray}
H=H^{diag} +H^{off}
=  \sum_{i=1}^N 
\bigg[ \vert q_{i+\frac{1}{2}}^{F} \rangle \langle q_{i+\frac{1}{2}}^{F} \vert 
+ \vert q_{i+\frac{1}{2}}^{AF} \rangle \langle q_{i+\frac{1}{2}}^{AF} \vert\bigg] 
\label{Htot2spinflip}
\end{eqnarray}
with the kets of Eq. \ref{qcpc}
\begin{eqnarray}
\vert q_{i+\frac{1}{2}}^{F} \rangle  && \equiv \sqrt{ D_{i+\frac{1}{2}}^{F}  }
\left( \frac{ \vert (+)_i (+)_{i+1} \rangle}{ e^{ \frac{\beta}{2} (h_i +h_{i+1})} } 
-  \frac{ \vert (-)_i (-)_{i+1} \rangle}{ e^{ - \frac{\beta}{2} (h_i +h_{i+1})} } \right) 
\nonumber \\
\vert q_{i+\frac{1}{2}}^{AF} \rangle  && \equiv \sqrt{ D_{i+\frac{1}{2}}^{AF}  }
\left( \frac{\vert (+)_i (-)_{i+1} \rangle}{ e^{ \frac{\beta}{2} (h_i -h_{i+1})} } 
-  \frac{ \vert (-)_i (+)_{i+1} \rangle}{ e^{  \frac{\beta}{2} (-h_i +h_{i+1})} }\right)
\label{qFAF}
\end{eqnarray}


\subsubsection{ Consequences for the explicit rate functions at level 2 and at level 2.1  }

For the Hamiltonian of Eq. \ref{Htot2spinflip},
the rate function $I^{[2]}  [ {\hat p}(.)] $ of Eq. \ref{I2densitymatrix} 
involving the empirical quantum density matrix $ {\hat \rho} = \vert \sqrt{ {\hat p}} \rangle  \langle \sqrt{ {\hat p}} \vert$ 
\begin{eqnarray}
I^{[2]}  [ {\hat p}(.)]
 = \langle \sqrt{ {\hat p}} \vert H \vert \sqrt{ {\hat p}} \rangle = \text{ Tr} (H {\hat \rho})
 = \sum_{i=1}^N 
\bigg[ \langle q_{i+\frac{1}{2}}^{F} \vert {\hat \rho}_{i,i+1}\vert q_{i+\frac{1}{2}}^{F} \rangle  
+\langle q_{i+\frac{1}{2}}^{AF} \vert  {\hat \rho}_{i,i+1} \vert q_{i+\frac{1}{2}}^{AF} \rangle \bigg] 
\label{Hmatrixelement2spinflip}
\end{eqnarray}
only involves the reduced quantum density matrices $ {\hat \rho}_{i,i+1} $
of size $4 \times 4$ associated to the two neighboring spins $(i,i+1)$
after taking the trace over all the other spins $S_{j\ne (i,i+1)}$
\begin{eqnarray}
 {\hat \rho}_{i,i+1} \equiv \text{ Tr}_{{j \ne (i,i+1)} } ( {\hat \rho} ) 
 = \sum_{S_1=\pm 1} ... \sum_{S_{i-1}=\pm 1}\sum_{S_{i+2}=\pm 1} ...\sum_{S_N=\pm 1} 
 \langle S_1,..,S_{i-1} ; S_{i+2},.. S_N \vert  {\hat \rho} \vert S_1,..,S_{i-1} ; S_{i+2},.. S_N \rangle
\label{rhoreduced2spinflip}
\end{eqnarray}
The real symmetric matrix elements read in terms of the empirical density ${\hat p}(.) $
\begin{eqnarray}
&& \langle S_i',S_{i+1}' \vert {\hat \rho}_{i,i+1} \vert S_i,S_{i+1} \rangle
 = \sum_{S_1=\pm 1} ... \sum_{S_{i-1}=\pm 1}\sum_{S_{i+2}=\pm 1} ...\sum_{S_N=\pm 1} 
 \langle S_1,..,S_{i-1},S_i',S_{i+1}', S_{i+2},.. S_N \vert  {\hat \rho} \vert S_1,..,S_{i-1} S_i',S_{i+1} S_{i+2},.. S_N \rangle
 \nonumber \\
&& = \sum_{S_1=\pm 1} ... \sum_{S_{i-1}=\pm 1}\sum_{S_{i+2}=\pm 1} ...\sum_{S_N=\pm 1}  
 \sqrt{ {\hat p}(S_1,..,S_{i-1},S_i',S_{i+1}', S_{i+2},.. S_N) {\hat p}(S_1,..,S_{i-1},S_i,S_{i+1}, S_{i+2},.. S_N)} 
\label{rhoreduced2spinflipoff}
\end{eqnarray}

Let us now compute separately the contributions of the diagonal and off-diagonal part
of the Hamiltonian in Eq. \ref{Hmatrixelement2spinflip}:

(i) the matrix element involving the diagonal part $H^{diag}$ of Eq. \ref{Hdiag2spinflip}
\begin{eqnarray}
 \langle \sqrt{ {\hat p}} \vert H^{diag} \vert \sqrt{ {\hat p}} \rangle
 =  \sum_{i=1}^N 
 \text{Tr}_{(i,i+1)}
&& \bigg(\bigg[ D_{i+\frac{1}{2}}^{F} 
\left( e^{ \beta (h_i +h_{i+1})}\pi_i^-  \pi_{i+1}^- +e^{-  \beta (h_i +h_{i+1})} \pi_i^+  \pi_{i+1}^+\right) 
\nonumber \\
&&+ D_{i+\frac{1}{2}}^{AF} \left(e^{ \beta (h_i -h_{i+1})} \pi_i^-  \pi_{i+1}^+ + e^{ \beta (-h_i +h_{i+1})} \pi_i^+  \pi_{i+1}^-
\right) \bigg]  {\hat \rho}_{i,i+1} \bigg)
\nonumber \\
 = \sum_{i=1}^N 
&& \bigg[ D_{i+\frac{1}{2}}^{F} 
\left( e^{ \beta (h_i +h_{i+1})} \langle -- \vert {\hat \rho}_{i,i+1} \vert -- \rangle
+e^{-  \beta (h_i +h_{i+1})}  \langle ++ \vert {\hat \rho}_{i,i+1} \vert ++ \rangle\right) 
\nonumber \\
&&+ D_{i+\frac{1}{2}}^{AF} \left(e^{ \beta (h_i -h_{i+1})}  \langle -+ \vert {\hat \rho}_{i,i+1} \vert -+ \rangle
+ e^{ \beta (-h_i +h_{i+1})} \langle +- \vert {\hat \rho}_{i,i+1} \vert +- \rangle
\right) \bigg]  
\label{Hmatrixelement2spinflipdiag}
\end{eqnarray}
only involves the diagonal elements $(S_i',S_{i+1}')=(S_i,S_{i+1})$ of the reduced density matrices $\rho_{i,i+1}$
of Eq. \ref{rhoreduced2spinflipoff}
\begin{eqnarray}
&& \langle S_i,S_{i+1} \vert {\hat \rho}_{i,i+1} \vert S_i,S_{i+1} \rangle
  = \sum_{S_1=\pm 1} ... \sum_{S_{i-1}=\pm 1}\sum_{S_{i+2}=\pm 1} ...\sum_{S_N=\pm 1}  
  {\hat p}(S_1,..,S_{i-1},S_i,S_{i+1}, S_{i+1},.. S_N)
\nonumber \\
&&  =  \sum_{S_1=\pm 1} ... \sum_{S_{i-1}=\pm 1}\sum_{S_{i+2}=\pm 1} ...\sum_{S_N=\pm 1}  
  \frac{1}{T} \int_0^T dt \prod_{n=1}^N \delta_{S_n(t) ,S_n} 
  = \frac{1}{T} \int_0^T dt \delta_{S_i(t) ,S_i}  \delta_{S_{i+1}(t) ,S_{i+1}} \equiv {\hat p}_{i,i+1}(S_i,S_{i+1})
\label{rhoreduced2spinflipdiag}
\end{eqnarray}
that represent the empirical local density ${\hat p}_{i,i+1}(S_i,S_{i+1})$ of the two spins on sites $(i,i+1)$ 
during the long trajectory over the time-window $[0,T]$ independently of the values of the other spins $j \ne (i,i+1)$,
with the normalization
\begin{eqnarray}
\text{Tr}_{(i,i+1)} ({\hat \rho}_{i,i+1} ) = \sum_{S_i=\pm} \sum_{S_{i+1} =\pm}\langle S_i,S_{i+1} \vert {\hat \rho}_{i,i+1} \vert S_i,S_{i+1} \rangle =\sum_{S_i=\pm} \sum_{S_{i+1} =\pm} {\hat p}_{i,i+1}(S_i,S_{i+1})
  = 1
\label{rhoreduced2spinflipdiagtrace}
\end{eqnarray}
So Eq. \ref{Hmatrixelement2spinflipdiag}
only involves the empirical two-spin densities ${\hat p}_{i,i+1}(S_i,S_{i+1})$
\begin{eqnarray}
 \langle \sqrt{ {\hat p}} \vert H^{diag} \vert \sqrt{ {\hat p}} \rangle
 = \sum_{i=1}^N 
&& \bigg[ D_{i+\frac{1}{2}}^{F} 
\left( e^{ \beta (h_i +h_{i+1})} {\hat p}_{i,i+1}(--)
+e^{-  \beta (h_i +h_{i+1})}  {\hat p}_{i,i+1}(++)\right) 
\nonumber \\
&&+ D_{i+\frac{1}{2}}^{AF} \left(e^{ \beta (h_i -h_{i+1})}  {\hat p}_{i,i+1}(-+)
+ e^{ \beta (-h_i +h_{i+1})} {\hat p}_{i,i+1}(+-)
\right) \bigg]  
\label{Hmatrixelement2spinflipdiaglocalp}
\end{eqnarray}

(ii) the matrix element involving the off-diagonal part $H^{off}$ of Eq. \ref{Hoff2spinflip}
\begin{eqnarray}
- \langle \sqrt{ {\hat p}} \vert H^{off} \vert \sqrt{ {\hat p}} \rangle
&& =  \sum_{i=1}^N \text{Tr}_{(i,i+1)}
\left( \left[ D_{i+\frac{1}{2}}^{F} \left(\sigma_i^+ \sigma_{i+1}^+ + \sigma_i^- \sigma_{i+1}^- \right) 
+ D_{i+\frac{1}{2}}^{AF} \left(\sigma_i^+ \sigma_{i+1}^- + \sigma_i^- \sigma_{i+1}^+ \right) \right]
 {\hat \rho}_{i,i+1}
\right)
\nonumber \\
&& =\sum_{i=1}^N 
 \bigg[ D_{i+\frac{1}{2}}^{F} \left( \langle ++ \vert {\hat \rho}_{i,i+1} \vert -- \rangle
 +  \langle --  \vert {\hat \rho}_{i,i+1} \vert ++ \rangle \right) 
+ D_{i+\frac{1}{2}}^{AF} \left( \langle +- \vert {\hat \rho}_{i,i+1} \vert-+ \rangle
+  \langle -+ \vert {\hat \rho}_{i,i+1} \vert+- \rangle\right) \bigg]
\nonumber \\
&& = \sum_{i=1}^N 
 \bigg[ 2 D_{i+\frac{1}{2}}^{F}  \langle ++ \vert {\hat \rho}_{i,i+1} \vert -- \rangle
 + 2 D_{i+\frac{1}{2}}^{AF} \langle +- \vert {\hat \rho}_{i,i+1} \vert-+ \rangle  \bigg]
\label{Hoffmatrixelement2spinflip}
\end{eqnarray}
only involves the off-diagonal elements $ \langle ++ \vert {\hat \rho}_{i,i+1} \vert -- \rangle =\langle --  \vert {\hat \rho}_{i,i+1} \vert ++ \rangle$ 
and $\langle +- \vert {\hat \rho}_{i,i+1} \vert-+ \rangle =\langle -+ \vert {\hat \rho}_{i,i+1} \vert+- \rangle$ of the 
symmetric reduced quantum density matrices ${\hat \rho}_{i,i+1} $ of Eq. \ref{rhoreduced2spinflip}.

In conclusion, the rate functions $I^{[2]}  [ {\hat p}(.)] $ and $ I^{[2.1]}[ {\hat p}(.) ;  {\hat A}] $ 
of Eqs \ref{legendreleftrightexplicitdiagetoff} and \ref{rate2.1jumpPAH}
can be written in terms of the two matrix elements $\langle \sqrt{ {\hat p}} \vert H^{diag} \vert \sqrt{ {\hat p}} \rangle $ and $\langle \sqrt{ {\hat p}} \vert H^{off} \vert \sqrt{ {\hat p}} \rangle $ of Eqs \ref{Hmatrixelement2spinflipdiaglocalp} and \ref{Hoffmatrixelement2spinflip}
that only contain matrix elements of the two-spin reduced density matrices ${\hat \rho}_{i,i+1} $.


\subsubsection{ Simplifications in the presence of translation-invariance for the spin chain  }

In the presence of translation invariance, all the fields $h_i$ parametrizing the steady state of Eq. \ref{steadyfactorized} coincide,
and all the coefficients $ D_{i+\frac{1}{2}}^{F,AF}$ of Eq. \ref{Hoff2spinflip}
are independent of the site $i$
\begin{eqnarray}
h_i && = h
\nonumber \\
D_{i+\frac{1}{2}}^{F} && = D^{F}
\nonumber \\
D_{i+\frac{1}{2}}^{AF} && = D^{AF}
\label{pure2spinflip}
\end{eqnarray}

Then the off-diagonal part $H^{off} $ of Eq. \ref{Hoff2spinflip} and the diagonal part $H^{diag} $ of Eq. \ref{Hdiag2spinflip}
 reduce to
\begin{eqnarray}
H^{off} && = - \sum_{i=1}^N \left[ D^{F} \left(\sigma_i^+ \sigma_{i+1}^+ + \sigma_i^- \sigma_{i+1}^- \right) 
+ D^{AF} \left(\sigma_i^+ \sigma_{i+1}^- + \sigma_i^- \sigma_{i+1}^+ \right)\right]
\nonumber \\
H^{diag} && =  \sum_{i=1}^N \left[ D^{F} 
\left( e^{ 2 \beta h}\pi_i^-  \pi_{i+1}^- +e^{- 2 \beta h} \pi_i^+  \pi_{i+1}^+\right) 
+ D^{AF} \left( \pi_i^-  \pi_{i+1}^+ +  \pi_i^+  \pi_{i+1}^- \right)\right] \nonumber \\
\label{Hdiag2spinflippure}
\end{eqnarray}
while the supersymmetric form of Eq. \ref{Htot2spinflip} for the Hamiltonian  now involves the kets 
\begin{eqnarray}
\vert q_{i+\frac{1}{2}}^{F} \rangle  && \equiv \sqrt{ D^{F}  }
\left( \frac{ \vert (+)_i (+)_{i+1} \rangle}{ e^{ \beta h} } 
-  \frac{ \vert (-)_i (-)_{i+1} \rangle}{ e^{ - \beta h} } \right) 
\nonumber \\
\vert q_{i+\frac{1}{2}}^{AF} \rangle  && \equiv \sqrt{ D^{AF}  }
\left( \vert (+)_i (-)_{i+1} \rangle 
-  \vert (-)_i (+)_{i+1} \rangle\right)
\label{qFAFpure}
\end{eqnarray}

As a consequence, the diagonal matrix element of Eq. \ref{Hmatrixelement2spinflipdiaglocalp}
\begin{eqnarray}
 \langle \sqrt{ {\hat p}} \vert H^{diag} \vert \sqrt{ {\hat p}} \rangle
&& = \sum_{i=1}^N 
 \bigg[ D^{F} 
\left( e^{ 2\beta h} {\hat p}_{i,i+1}(--)
+e^{-  2 \beta h}  {\hat p}_{i,i+1}(++)\right) 
+ D^{AF} \left(  {\hat p}_{i,i+1}(-+)
+  {\hat p}_{i,i+1}(+-)
\right) \bigg]  
\nonumber \\
&& = N \bigg[ D^{F} 
\left( e^{ 2\beta h} {\hat p}^{av}_2(--)
+e^{-  2 \beta h}  {\hat p}^{av}_2\right) 
+ D^{AF} \left(  {\hat p}^{av}_2
+  {\hat p}^{av}_2
\right) \bigg]  
\label{Hmatrixelement2spinflipdiaglocalpure}
\end{eqnarray}
only involves the spatial-average over the $N$ sites of the empirical two-spin densities ${\hat p}_{i,i+1}(S, S') $
\begin{eqnarray}
{\hat p}^{av}_2(S ,S') \equiv  \frac{1}{N} \sum_{i=1}^N  {\hat p}_{i,i+1}(S, S')
\label{p2spinspatialav}
\end{eqnarray}
while the off-diagonal matrix element Eq. \ref{Hoffmatrixelement2spinflip}
\begin{eqnarray}
- \langle \sqrt{ {\hat p}} \vert H^{off} \vert \sqrt{ {\hat p}} \rangle
&& = \sum_{i=1}^N 
 \bigg[ 2 D^{F}  \langle ++ \vert {\hat \rho}_{i,i+1} \vert -- \rangle
 + 2 D^{AF} \langle +- \vert {\hat \rho}_{i,i+1} \vert-+ \rangle  \bigg]
 \nonumber \\
&& = N 
 \bigg[ 2 D^{F}  \langle ++ \vert {\hat \rho}_2^{av} \vert -- \rangle
 + 2 D^{AF} \langle +- \vert {\hat \rho}_2^{av} \vert-+ \rangle  \bigg]
\label{Hoffmatrixelement2spinflippure}
\end{eqnarray}
only involves the spatial-averages of the two-spin-flip matrix elements of the density matrices ${\hat \rho}_{i,i+1}$
\begin{eqnarray}
\langle -S,-S' \vert {\hat \rho}^{av}_2 \vert S,S' \rangle \equiv  \frac{1}{N} \sum_{i=1}^N  \langle -S,-S' \vert {\hat \rho}_{i,i+1} \vert S,S' \rangle
\label{rhooff2spinspatialav}
\end{eqnarray}

Finally, the rate function $I^{[2]}  [ {\hat p}(.)] $ of Eq. \ref{Hmatrixelement2spinflip}
based on the supersymmetric Hamiltonian $H$ 
with the kets of Eq. \ref{qFAFpure} 
becomes
\begin{eqnarray}
I^{[2]}  [ {\hat p}(.)]
&&  = N D^F  \left( \frac{ \langle ++ \vert}{ e^{ \beta h} } -  \frac{ \langle -- \vert}{ e^{ - \beta h} } \right)  
 {\hat \rho}^{av}_2
\left( \frac{ \vert ++ \rangle}{ e^{ \beta h} } -  \frac{ \vert -- \rangle}{ e^{ - \beta h} } \right)  
\nonumber \\
&& + N D^{AF} 
\bigg( \langle +- \vert -  \langle -+ \vert \bigg)   {\hat \rho}^{av}_2 
\bigg( \vert +- \rangle -  \vert -+ \rangle\bigg)
\label{Hmatrixelement2spinflippure}
\end{eqnarray}
in terms of the spatial-average ${\hat \rho}^{av}_2 $ of the density matrices ${\hat \rho}_{i,i+1}$
\begin{eqnarray}
 {\hat \rho}^{av}_2 \ \equiv  \frac{1}{N} \sum_{i=1}^N   {\hat \rho}_{i,i+1} 
\label{rhotot2spinspatialav}
\end{eqnarray}


\subsection{ Single-spin-flip dynamics of spin chains with Ising steady states  }

\label{sec_1spinflip}

In this subsection, we consider a spin chain of $N$ spins 
where the steady state is given by the Boltzmann distribution
associated to the Ising energy containing fields $h_i$ on sites $i=1,..,N$
and couplings $J_{i+\frac{1}{2}}$ between nearest-neighboring spins $(S_i,S_{i+1})$
\begin{eqnarray}
P_*(C=\{S_1,S_2,..,S_N\}) =  \frac{ e^{ \displaystyle  \beta \sum_{i=1}^N \left( h_i S_i+ J_{i+\frac{1}{2}} S_i S_{i+1} \right) } }{ Z_N }
\label{steadyspinchain}
\end{eqnarray}
while the partition function $Z_N$ ensures the normalization over the $2^N$ configurations.


\subsubsection{ Supersymmetric quantum Hamiltonian associated to single-spin-flip dynamics  }

In this section, we focus on the single-spin-flip dynamics $S_i \rightarrow -S_i$ described by the Pauli matrix $\sigma_i^x=\sigma_i^+ + \sigma_i^-$,
while the other $(N-1)$ spins $S_{j \ne i}$ remain unchanged.
So the off-diagonal part of the Hamiltonian of Eq. \ref{offdiag}
can be written as
\begin{eqnarray}
H^{off} = - \sum_{i=1}^N \left[ D_i^{++}   \pi_{i-1}^+ \sigma_i^x\pi_{i+1}^+
+D_i^{--}   \pi_{i-1}^- \sigma_i^x \pi_{i+1}^-
+D_i^{+-}   \pi_{i-1}^+ \sigma_i^x \pi_{i+1}^-
+D_i^{-+}   \pi_{i-1}^- \sigma_i^x \pi_{i+1}^+
\right] 
\label{Hoff1spinflip}
\end{eqnarray}
where the four coefficients $D_i^{S_{i-1}=\pm,S_{i+1}=\pm} \geq 0$ parametrize the single-spin-flip rates of $S_i$
as a function of the values of two neighboring spins $(S_{i-1},S_{i+1} )$, while $\pi^{\pm}_j$ are the projectors introduced in Eq. \ref{pauliprojectors}.  

The diagonal part of Eq. \ref{diag}
 then reads using the steady state of Eq. \ref{steadyspinchain}
\begin{eqnarray}
H^{diag} = \sum_{i=1}^N \sum_{S=\pm} \sum_{S'=\pm} 
 D_i^{SS'}   \pi_{i-1}^S
  \left( e^{ - \beta (J_{i-\frac{1}{2}} S+h_i +J_{i+\frac{1}{2}} S')} \pi_i^+ 
 + e^{  \beta (J_{i-\frac{1}{2}} S+h_i +J_{i+\frac{1}{2}} S')} \pi_i^- \right) 
 \pi_{i+1}^{S' }
\label{Hdiag1spinflip}
\end{eqnarray}

Finally, the full Hamiltonian $H$ can be written in the supersymmetric form of Eqs \ref{Hsum} \ref{Hlinkproj}
\begin{eqnarray}
H=H^{diag} +H^{off}
=  \sum_{i=1}^N \sum_{S=\pm} \sum_{S'=\pm}  \vert q_i^{SS'} \rangle \langle q_i^{SS'} \vert 
\label{Htot1spinflip}
\end{eqnarray}
with the kets of Eq. \ref{qcpc}
\begin{eqnarray}
\vert q_i^{SS'} \rangle  && \equiv \sqrt{ D_i^{SS'}  }
\left( \frac{ \vert S_{i-1}=S; S_i=+; S_{i+1}=S' \rangle}{ e^{ \frac{\beta}{2} (J_{i-\frac{1}{2}} S+h_i +J_{i+\frac{1}{2}} S')} } 
-  \frac{ \vert S_{i-1}=S; S_i=-; S_{i+1}=S' \rangle}{ e^{ - \frac{\beta}{2} (J_{i-\frac{1}{2}} S+h_i +J_{i+\frac{1}{2}} S')} } \right) 
\label{q1spinflip}
\end{eqnarray}


\subsubsection{ Consequences for the explicit rate functions at level 2 and at level 2.1  }

So the difference with the the subsection \ref{sec_1spinflip} is that 
the rate functions $I^{[2]}  [ {\hat p}(.)] $ and $ I^{[2.1]}[ {\hat p}(.) ;  {\hat A}] $ 
of Eqs \ref{legendreleftrightexplicitdiagetoff} and \ref{rate2.1jumpPAH}
 written in terms of the two matrix elements $\langle \sqrt{ {\hat p}} \vert H^{diag} \vert \sqrt{ {\hat p}} \rangle $ and $\langle \sqrt{ {\hat p}} \vert H^{off} \vert \sqrt{ {\hat p}} \rangle $ of Eqs \ref{Hmatrixelement2spinflipdiaglocalp} and \ref{Hoffmatrixelement2spinflip}
will now involve matrix elements of the three-spin reduced density matrices ${\hat \rho}_{i-1,i,i+1} $
 \begin{eqnarray}
 {\hat \rho}_{i-1,i,i+1} && \equiv \text{ Tr}_{{j \ne (i-1,i,i+1)} } ( {\hat \rho} ) 
\nonumber \\
&&  = \sum_{S_1=\pm 1} ... \sum_{S_{i-2}=\pm 1}\sum_{S_{i+2}=\pm 1} ...\sum_{S_N=\pm 1} 
 \langle S_1,..,S_{i-2} ; S_{i+2},.. S_N \vert  {\hat \rho} \vert S_1,..,S_{i-2} ; S_{i+2},.. S_N \rangle
\label{rhoreduced3spinflip}
\end{eqnarray}
instead of the two -spin reduced density matrices ${\hat \rho}_{i,i+1} $ of Eq. \ref{rhoreduced2spinflip}.


\subsubsection{ Simplifications in the presence of translation-invariance for the spin chain  }

In the presence of translation invariance of the parameters
\begin{eqnarray}
h_i && = h
\nonumber \\
D_i^{SS'} && = D^{SS'}
\label{pure1spinflip}
\end{eqnarray}
the matrix elements $\langle \sqrt{ {\hat p}} \vert H^{diag} \vert \sqrt{ {\hat p}} \rangle $ and $\langle \sqrt{ {\hat p}} \vert H^{off} \vert \sqrt{ {\hat p}} \rangle $ and thus the rate functions
$I^{[2]}  [ {\hat p}(.)] $ and $ I^{[2.1]}[ {\hat p}(.) ;  {\hat A}] $
will now involve the the spatial-average $ {\hat \rho}^{av}_3 $ of the three-spin density matrices ${\hat \rho}_{i-1,i,i+1}$ of Eq. \ref{rhoreduced3spinflip}
\begin{eqnarray}
 {\hat \rho}^{av}_3 \ \equiv  \frac{1}{N} \sum_{i=1}^N   {\hat \rho}_{i-1,i,i+1} 
\label{rhotot3spinspatialav}
\end{eqnarray}
instead of Eq. \ref{rhotot2spinspatialav}.


\subsection{ Discussion  }

As shown by these two examples of inhomogeneous chains of $N$ spins with ${\cal N}= 2^N$ configurations,
where the transitions rates are local in space and involve only two or three neighboring spins,
the density matrix ${\hat \rho} $ of size $ {\cal N} \times {\cal N}$ actually appears in rate functions
only via its $N$ reduced density matrices ${\hat \rho}_{i,i+1} $  or ${\hat \rho}_{i-1,i,i+1} $ involving only two or three neighboring spins.
In the presence of translation invariance, the simplifications are even greater,
since only the spatial-average $ {\hat \rho}^{av}_2 $ 
or $ {\hat \rho}^{av}_3 $ actually appears in rate functions.

These simplifications of rate functions in terms of reduced density matrices
can be adapted for other reversible dynamics of pure or inhomogeneous many-body models with local transition rates.


\section{ Conclusions  }

\label{sec_conclusions}

In the main text, we have revisited the explicit large deviations at various levels of reversible Markov jump processes from the perspective of the quantum Hamiltonian $H$ that can be obtained from the Markov generator via a similarity transformation, whose supersymmetric properties were recalled in subsection \ref{subsec_susy}.

We have first focused on the large deviations at level 2 concerning the distribution of the empirical density ${\hat p}(C) $ 
seen during a large-time window $[0,T]$
in order to rewrite the explicit Donsker-Varadhan rate function at level 2 for a reversible Markov jump process
as the matrix element $I^{[2]}[{\hat p}(.) ] = \langle \sqrt{ {\hat p} } \vert H \vert \sqrt{{\hat p}} \rangle $ 
of Eq. \ref{legendreleftrightexplicit} that involves only the quantum supersymmetric Hamiltonian $H$ and the square-root ket $\vert \sqrt{ {\hat p}} \rangle $,
with the consequences of Eqs  \ref{normaweightswn} \ref{I2assumofsquares} \ref{I2densitymatrix}.
We have also discussed the corresponding explicit 
generator $w^{Cond[\hat p(.)]} $ conditioned to produce the given empirical density ${\hat p}(.) $ in Eq. \ref{wconditionedp}.

We have then recalled the explicit level 2.5 concerning the joint distribution of the empirical density ${\hat p}(C) $ and of the empirical transitions between configurations in order to discuss various intermediate levels between 2.5 and 2 whose rate functions are still explicit for reversible Markov jump processes, namely :

(i) the joint distribution ${\cal P}^{[2.25']}_T[ {\hat p}(.);  {\hat j}(. , .)]  $ of the empirical density ${\hat p}(.)$ and the empirical currents ${\hat j}(.,.)$
in Eqs \ref{rate2.25jumpPj} \ref{proba2.25jumpPj}, with the corresponding conditioned generator $w^{Cond[{\hat p}(.);  {\hat j}(. , .)]} $ of Eq. \ref{wconditionedpj};

(ii) the joint distribution ${\cal P}^{[2.25]}_T[ {\hat p}(.);  {\hat a}(. , .)]  $  of the empirical density ${\hat p}(.)$ and the empirical activities ${\hat a}(.,.)$ in Eqs \ref{rate2.25jumpPa} \ref{proba2.25jumpPa}, 
with the corresponding conditioned generator $w^{Cond[{\hat p}(.);  {\hat a}(. , .)]} $  of Eq. \ref{wconditionedpa};

(iii) the joint distribution ${\cal P}^{[2.1]}_T[ {\hat p}(.);  {\hat A}]  $ of the empirical density ${\hat p}(.)$ and the total empirical activity ${\hat A}(.,.)$ of Eq. \ref{proba2.1jumpPAtot} \ref{rate2.1jumpPA} \ref{rate2.1jumpPAdifference},
where the rate function $ I^{[2.1]}[ {\hat p}(.) ;  {\hat A}] $ was rewritten in Eq. \ref{rate2.1jumpPAH}
in terms of the two matrix elements $ \langle \sqrt{ {\hat p} } \vert H \vert \sqrt{{\hat p}} \rangle$ and $\langle \sqrt{ {\hat p} } \vert H^{off} \vert \sqrt{{\hat p}} \rangle $, 
with the corresponding expression of Eq. \ref{geneAproba2.1H}
for the generating function ${\cal Z}^{[2.1]}_T[ {\hat p}(.); \lambda ] $ of the total empirical acitivity ${\hat A}(.,.) $
at fixed empirical density ${\hat p}(.) $,
while the corresponding condition generator $w^{Cond[{\hat p}(.);  {\hat A}]} $ was written in Eq. \ref{wconditionedpatot}.

Finally, this general framework was applied to pure or random spin chains with single-spin-flip or two-spin-flip transition rates, where the supersymmetric Hamiltonian $H$ correspond to quantum spin chains with local interactions involving Pauli matrices of two or three neighboring sites, in order to show how various rate functions can be rewritten in terms of reduced density matrices involving only two or three neighboring sites.

In Appendix \ref{app_diffusion}, 
we have revisited the large deviations properties of reversible diffusion processes
from the perspective of quantum supersymmetric Hamiltonians in continuous space,
in order to stress the similarities and the differences with Markov jump processes
in the last subsection \ref{subapp_differences}.

Since the main goal of the present paper was to stress the importance of supersymmetric quantum Hamiltonians 
to analyze reversible Markov processes, it is natural to ask what could be the appropriate generalizations for irreversible processes. For systems that are out-of-equilibrium only as a consequence of the boundary-driving, the similarity transformation towards supersymmetric quantum Hamiltonians is still very useful 
as discussed for one-dimensional rings \cite{c_lyapunov} or for one-dimensional intervals in contact with two reservoirs \cite{c_boundarydriven}. For systems where detailed-balance is broken by the dynamical rules in the bulk, there are two possibilities :

(a) one can still make the same similarity transformation as in Eqs \ref{relationPpsi} or \ref{ppsi} in order to analyze the properties of the non-hermitian quantum Hamiltonians that emerge, as discussed in \cite{us_gyrator} for the case of irreversible diffusion processes in arbitrary dimension; 

(b) or one can analyze directly the non-hermitian quantum Hamiltonian corresponding to the opposite of the generator
with its non-hermitian supersymmetric factorization $H=-w= {\bold I}^{\dagger} {\bold J}$
 involving the current operator ${\bold J} $ and the divergence ${\bold I}^{\dagger} $ appearing in the continuity equation, as discussed both for Markov jump processes and for diffusions in \cite{c_susynonhermitiannoneq}.


\appendix

\section{ Comparison with explicit large deviations for reversible diffusion processes  }

\label{app_diffusion}

In this Appendix, we revisited the large deviations properties of reversible diffusion processes
from the perspective of quantum supersymmetric Hamiltonians in continuous space. 
Indeed, the comparison between these two types of continuous-time Markov processes
is often very useful to better understand each of them.

\subsection{ Reversible Fokker-Planck dynamics and the corresponding supersymmetric quantum Hamiltonians}

The Fokker-Planck equation for the probability density $P_t(\vec r)$ to be at position $\vec r$ at time $t$
can be written as the continuity equation (analog of Eq. \ref{mastercontinuity} of the main text)
\begin{eqnarray}
 \partial_t P_t( \vec r )    = -  \vec \nabla . \vec J_t( \vec r ) 
\label{fokkerplanck}
\end{eqnarray}
where the current
\begin{eqnarray}
\vec J_t( \vec r )    =   P_t(\vec r) \vec F (\vec r) - D(\vec r)  \vec \nabla P_t(\vec r)   
\label{fokkerplanckcurrent}
\end{eqnarray}
involves the force $\vec F (\vec r) $ and the diffusion coefficient $D(\vec r)$.

When the dynamics converging towards the steady state $P_*(C)$
satisfies detailed-balance, the associated steady current should vanish
\begin{eqnarray}
0 = \vec J_*( \vec r )    =   P_*(\vec r) \vec F (\vec r) - D(\vec r)  \vec \nabla P_*(\vec r) 
\label{DBdiff}
\end{eqnarray}
so that the force can be replaced by
\begin{eqnarray}
 \vec F (\vec r) = D(\vec r)  \vec \nabla \ln P_*(\vec r)
\label{forceDB}
\end{eqnarray}
which is the analog of the parametrization of Eq. \ref{detailedratio}
for the detailed-balance transition rates of the main text.
The Fokker-Planck dynamics of
Eq. \ref{fokkerplanck} then becomes
\begin{eqnarray}
 \partial_t P_t( \vec r )    =  
 -  \vec \nabla .  \left[ D(\vec r) \left(   [ \vec \nabla \ln P_*(\vec r) ] -  \vec \nabla    \right) P_t(\vec r) \right]
\label{fokkerplanckrev}
\end{eqnarray}

As explained in textbooks \cite{gardiner,vankampen,risken} and in specific applications to various models (see for instance \cite{glauber,Felderhof,siggia,kimball,peschel,jpb_antoine,pierre,texier,us_eigenvaluemethod,Castelnovo,c_pearson,c_boundarydriven}), generators of reversible Markov processes are related to quantum hermitian Hamiltonians
via similarity transformations.
For diffusion processes,
the standard change of variables analogous to Eq. \ref{relationPpsi}
\begin{eqnarray}
P_t(\vec r) = \sqrt{ P_*(\vec r)}  \psi_t(\vec r)
\label{ppsi}
\end{eqnarray}
transforms the Fokker-Planck Eq. \ref{fokkerplanckrev}
 into the euclidean Schr\"odinger equation
 \begin{eqnarray}
\partial_t \psi_t ( \vec r ) = - H \psi_t ( \vec r )
\label{schrodinger}
\end{eqnarray}
where the quantum Hamiltonian obtained from the Fokker-Planck generator of Eq. \ref{fokkerplanckrev}
via a similarity transformation analog to Eq. \ref{Hsimilarity}
\begin{eqnarray}
H && = \frac{1}{\sqrt{ P_*(\vec r)}}    \vec \nabla .  \left[ D(\vec r) \left(   [ \vec \nabla \ln P_*(\vec r) ] -  \vec \nabla    \right) \sqrt{ P_*(\vec r)} \right]
\nonumber \\
&&=\left(   \frac{ [ \vec \nabla \ln P_*(\vec r) ]}{2}  +  \vec \nabla    \right) D(\vec r) 
\left(   \frac{ [ \vec \nabla \ln P_*(\vec r) ]}{2}  -  \vec \nabla    \right)
 \equiv  \vec Q^{\dagger} . \vec Q
\label{hsusy}
\end{eqnarray}
is supersymmetric in terms of the two first-order operators 
(see the review \cite{review_susyquantum} on supersymmetric quantum mechanics in continuous space)
\begin{eqnarray}
\vec Q  && \equiv    \sqrt{ D(\vec r) }  \left(   \frac{ [ \vec \nabla \ln P_*(\vec r) ]}{2}  -  \vec \nabla    \right)
\nonumber \\
\vec Q^{\dagger}  &&\equiv  \left(   \frac{ [ \vec \nabla \ln P_*(\vec r) ]}{2}  +  \vec \nabla    \right) \sqrt{ D(\vec r ) }
\label{qsusy}
\end{eqnarray}
This supersymmetric factorization is very standard for diffusion processes in continuous space (see for instance  \cite{jpb_antoine,pierre,texier,c_lyapunov,us_gyrator,c_pearson,c_boundarydriven}).

In particular, the ground state at zero energy of the Hamiltonian $H$ 
\begin{eqnarray}
  \psi_0 (\vec r) = \sqrt{ P_*(\vec r) }
\label{psi0r}
\end{eqnarray}
is annihilated by the operator $\vec Q$ 
\begin{eqnarray}
\vec Q   \psi_0 (\vec r) = 0
\label{psi0rannihilate}
\end{eqnarray}
which is the analog of Eq. \ref{Hlinkannihilation}.
It is convenient to replace the steady state $P_*(\vec r) = \psi_0^2 (\vec r) $
into Eqs \ref{qsusy} to obtain the first order operators in terms of the ground state $\psi_0 (\vec r) $
\begin{eqnarray}
\vec Q  && \equiv    \sqrt{ D(\vec r) }  \left(    [ \vec \nabla \ln \psi_0(\vec r) ]  -  \vec \nabla    \right)
\nonumber \\
\vec Q^{\dagger}  &&\equiv  \left(    [ \vec \nabla \ln \psi_0(\vec r) ]  +  \vec \nabla    \right) \sqrt{ D(\vec r ) }
\label{qsusypsi}
\end{eqnarray}

As in Eq. \ref{propagator}, the Hamiltonian $H$ governs the Fokker-Planck propagator $P_t( \vec r \vert \vec r_0 ) $
via its spectral decomposition that involves here either an infinite number of discrete energies or 
a finite number of discrete energies before a continuum spectrum.


\subsection{ Explicit large deviations at Level 2.5 for the empirical density and current of arbitrary diffusions   }

For a diffusive trajectory $\vec r(0 \leq t \leq T) )$ over the large-time window $t \in [0,T]$,
the empirical density measuring the faction of time spent around at each position $\vec r$
\begin{eqnarray}
{\hat P}(\vec r ) \equiv \frac{1}{T} \int_0^T dt \delta( \vec r(t)- \vec r)
\label{Pempi}
\end{eqnarray}
is normalized to unity
\begin{eqnarray}
\int d^d \vec r {\hat P}(\vec r ) =1
\label{Pempinorma}
\end{eqnarray}
while the empirical current
\begin{eqnarray}
{\hat {\vec J}}(\vec r ) \equiv \frac{1}{T} \int_0^T dt \frac{d \vec r(t) }{dt} \delta( \vec r(t)- \vec r)
\label{Jempi}
\end{eqnarray}
should be divergenceless 
\begin{eqnarray}
\vec \nabla . {\hat {\vec J}}(\vec r ) =0
\label{divJempi}
\end{eqnarray}

For arbitrary diffusion processes governed by the Fokker-Planck dynamics of Eqs \ref{fokkerplanck} \ref{fokkerplanckcurrent}, 
the joint distribution ${\cal P}^{[2.5]}_T[ {\hat P}(.), {\hat {\vec J}}(.)] $ of the empirical density ${\hat P}(.) $
and the empirical current ${\hat {\vec J}}(.) $
satisfying the constitutive constraints of Eqs \ref{Pempinorma} \ref{divJempi}
follows the large deviation form for large $T$
\cite{wynants_thesis,maes_diffusion,chetrite_formal,engel,chetrite_HDR,c_lyapunov,c_inference,c_susyboundarydriven,c_missing,c_SmallNoise}
\begin{eqnarray}
 {\cal P}^{[2.5]}_T[ {\hat P}(.), {\hat {\vec J}}(.)]   
 \opsimeq_{T \to +\infty}  \delta \left(\int d^d \vec r {\hat P}(\vec r) -1  \right)
\left[ \prod_{\vec r }  \delta \left(  \vec \nabla . {\hat {\vec J}}(\vec r) \right) \right] 
e^{- \displaystyle I^{[2.5]}[ {\hat P}(.), {\hat {\vec J}}(.)]
 }
\label{ld2.5diff}
\end{eqnarray}
with the explicit rate function 
\begin{eqnarray}
 I^{[2.5]}[ {\hat P}(.), {\hat {\vec J}}(.)]
 =
\int \frac{d^d \vec r}{ 4 D (\vec r) {\hat P}(\vec r) } \left[ {\hat {\vec J}}(\vec r) - {\hat P}(\vec r)  \vec F (\vec r)+D (\vec r) \vec \nabla {\hat P}(\vec r) \right]^2
\label{rate2.5diff}
\end{eqnarray}

For any given empirical density $ {\hat P}(.)$,
it is interesting to consider the reversal of all the empirical currents ${\hat J} (\vec r) \to - {\hat J} (\vec r)$:
the divergenceless constraint is still satisfied,
while the difference between the two rate functions reduce to the linear contribution with respect to the empirical current
${\hat {\vec J}}(\vec r) $
\begin{eqnarray}
 I^{[2.5]}[ {\hat P}(.), {\hat {\vec J}}(.)] -  I^{[2.5]}[ {\hat P}(.), - {\hat {\vec J}}(.)]
 =
\int d^d \vec r {\hat {\vec J}}(\vec r) . \left[  -   \frac{ \vec F (\vec r)}{D (\vec r) } + \vec \nabla \ln {\hat P}(\vec r) \right]
\label{rate2.5difference}
\end{eqnarray}
This property is the analog of Eq. \ref{rate2.5jumpPajdifference} concerning Markov jump processes
and can be considered as an example of the Gallavotti-Cohen fluctuation relations
(see \cite{galla,kurchan_langevin,Leb_spo,maes1999,jepps,derrida-lecture,harris_Schu,kurchan,searles,zia,chetrite_thesis,maes2009,maes2017,chetrite_HDR} and references therein).


\subsection{ Explicit large deviations at Level 2 for the empirical density of reversible diffusion processes  }

When the force $\vec F (\vec r) $  satisfies the detailed-balance condition of Eq. \ref{forceDB}, 
the difference of Eq. \ref{rate2.5difference} vanishes
via integration by parts using the divergenceless property of Eq. \ref{divJempi}
for the empirical current ${\hat {\vec J}}(.) $
\begin{eqnarray}
 I^{[2.5]}[ {\hat P}(.), {\hat {\vec J}}(.)] -  I^{[2.5]}[ {\hat P}(.), - {\hat {\vec J}}(.)]
&& =
\int d^d \vec r {\hat {\vec J}}(\vec r) . \left[   \vec \nabla \ln \frac{ {\hat P}(\vec r) }{ {\hat P}_*(\vec r)}  \right]
= - \int d^d \vec r  . \left[    \ln \left( \frac{ {\hat P}(\vec r) }{ {\hat P}_*(\vec r)} \right) \right] \vec \nabla .{\hat {\vec J}}(\vec r)
=0
\label{rate2.5differenceDB}
\end{eqnarray}
One can still be interested into the large deviations involving non-vanishing empirical currents ${\hat {\vec J}}(.) \ne \vec 0 $
as described by the explicit level 2.5 of Eq. \ref{ld2.5diff},
but one can also consider the vanishing optimal empirical currents ${\hat {\vec J}}^{opt}(.) = \vec 0 $
to obtain the explicit large deviations at level 2 for the distribution of the empirical density
\begin{eqnarray}
 P^{[2]}_T[ {\hat P}(.)]   \opsimeq_{T \to +\infty}  \delta \left(\int d^d \vec r {\hat P}(\vec r) -1  \right)
e^{- \displaystyle I^{[2]}[ {\hat P}(.)]
 }
\label{ld2diff}
\end{eqnarray}
where the rate function at Level 2 is obtained from the rate function at Level 2.5 of Eq. \ref{rate2.5diff}
for vanishing current ${\hat {\vec J}}^{opt}(.) = \vec 0 $
\begin{eqnarray}
I^{[2]}[ {\hat P}(.)] =
   I^{[2.5]}[ {\hat P}(.), {\hat {\vec J}}^{opt}(.) = \vec 0]
&& =\frac{1}{4} \int d^d \vec r \  D (\vec r) {\hat P}(\vec r)  
  \left[       \vec \nabla \ln( {\hat P}(\vec r) ) - \vec \nabla \ln( {\hat P}_*(\vec r) )\right]^2 
  \nonumber \\
&&  = \frac{1}{4} \int d^d \vec r \  D (\vec r) {\hat P}(\vec r)  
  \left[       \vec \nabla \ln \left( \frac{ {\hat P}(\vec r) }{ {\hat P}_*(\vec r) } \right) \right]^2 
\label{rate2diff}
\end{eqnarray}

To make the link with the supersymmetric quantum Hamiltonian,
it is convenient to replace the steady state $P_*(\vec r) = \psi_0^2 (\vec r) $
 by the square of the groundstate of Eq. \ref{psi0r},
 and to replace similarly the empirical density ${\hat P}(\vec r) =  {\hat \psi}^2(\vec r)$
to rewrite the rate function of Eq. \ref{rate2diff}
\begin{eqnarray}
I^{[2]}[ {\hat P}(.)={\hat \psi}^2(\vec r)] 
&& = \int d^d \vec r \  D (\vec r) {\hat \psi}^2(\vec r)  
  \left[       \vec \nabla \ln( {\hat \psi}(\vec r) ) - \vec \nabla \ln( {\hat \psi}_0(\vec r) )\right]^2 
   \nonumber \\
  && = \int d^d \vec r \  D (\vec r)  
  \left[       \vec \nabla ( {\hat \psi}(\vec r)  - {\hat \psi}(\vec r) \vec \nabla \ln( {\hat \psi}_0(\vec r) )\right]^2 
    \nonumber \\
  && = \int d^d \vec r \   
  \left[   \vec Q    {\hat \psi}(\vec r)  \right]^2 
\label{rate2diffsusycalcul}
\end{eqnarray}
where one recognizes the first-order operator $\vec Q$ of Eq. \ref{qsusypsi} applied to ${\hat \psi}(\vec r) 
= \sqrt{{\hat P}(\vec r) } $.
As a consequence, the rate function at level 2 can be rewritten  
\begin{eqnarray}
I^{[2]}[ {\hat P}(.) = \int d^d \vec r \   
  \left[   \vec Q    \sqrt{{\hat P}(\vec r) }  \right]^2 
  = 
   \int d^d \vec r \left( \sqrt{{\hat P}(\vec r) } \vec Q^{\dagger} . \vec Q \sqrt{{\hat P}(\vec r) } \right)  
 \equiv \langle \sqrt{{\hat P}} \vert H \vert \sqrt{{\hat P} }\rangle
\label{rate2diffsusy}
\end{eqnarray}
as the matrix element $\langle \sqrt{{\hat P}} \vert H \vert \sqrt{{\hat P} }\rangle $ 
involving the supersymmetric Hamiltonian $H=\vec Q^{\dagger} . \vec Q$,
which is the analog of Eq. \ref{legendreleftrightexplicit} concerning Markov jump processes.

 
\subsection{ Explicit conditioned generator of a reversible diffusion with respect to the empirical density ${\hat P}(.)$ } 

As in the subsection \ref{subsec_conditioning} of the main text concerning Markov jump processes, 
the conditioned generator of a reversible diffusion with respect to the empirical density ${\hat P}(.)$
is explicit : the diffusion coefficient $D(\vec r)$ cannot change,
so the conditioned Fokker-Planck generator only contains a different force $ \vec F^{Cond} (\vec r) $
with respect to the initial generator.
Since this conditioned generator should satisfy detailed-balance with respect to its conditioned steady state 
$P_*^{Cond} (\vec r)= {\hat P}(\vec r)$,
the conditioned force reduces to 
\begin{eqnarray}
  \vec F^{Cond} (\vec r) = D(\vec r)  \vec \nabla \ln {\hat P}(\vec r)
\label{forceDBcond}
\end{eqnarray}
instead of the initial force $ \vec F (\vec r) $ of Eq. \ref{forceDB}.


\subsection{ Discussion}

\label{subapp_differences}

In summary, many ideas are very similar with the case of Markov jump processes discussed in the main text:

(a) reversible generators can be mapped onto supersymmetric Hamiltonians $H$ via similarity transformations; 

(b) these supersymmetric Hamiltonians are useful to rewrite the explicit rate function at level 2
as Eq. \ref{rate2diffsusy} which is the direct analog of Eq. \ref{legendreleftrightexplicit}.
 
 (c) the conditioning of a reversible generator towards any given empirical density is explicit.

 There are however some important differences:

(i) in diffusion processes, there are no fluctuating empirical activities, 
but only the fluctuating empirical density and fluctuating empirical currents 
 (see more detailed discussions in \cite{c_missing}) :
so the standard level 2.5 for diffusion processes concerning their joint distribution ${\cal P}^{[2.5]}_T[ {\hat P}(.), {\hat {\vec J}}(.)]   $ of Eq. \ref{ld2.5diff} is actually more the analog of the level $2.25'$ of Markov jump processes
concerning ${\cal P}^{[2.25'}_T[ {\hat p}(.);  {\hat j}(. , .)]  $  
of Eq. \ref{proba2.25jumpPj}, while the levels 2.5, 2.25 and 2.1 that contain some empirical activities
of Markov processes as discussed in the main text
have no counterparts for diffusion processes.

(ii) in diffusion processes, the diffusion coefficient $D(\vec r)$ cannot fluctuate (see more detailed discussions in \cite{c_inference}), in contrast to the empirical coefficients ${\hat D}$ of Eq. \ref{detailedratioempi}
discussed in the main text for Markov jump processes.


\end{document}